\documentclass[referee,a4paper,12pt,traditabstract]{jswsc} 

\usepackage{amssymb}
\usepackage{booktabs}
\usepackage{multirow}
\usepackage{graphicx}
\usepackage{txfonts}
\usepackage{subfigure}
\usepackage{epstopdf}
\usepackage[displaymath,mathlines]{lineno}
\usepackage[authoryear,round]{natbib}
\usepackage[backref]{hyperref}
\usepackage{url}

\usepackage{todonotes}

\hypersetup{colorlinks=true,citecolor=cyan,urlcolor=cyan,linkcolor=blue}

\usepackage{xspace}

\begin{document}


   \title{Comprehensive solar eruption analyses enabled by the tools of the SOLER project}

   
   \titlerunning{SOLER tools}

   \authorrunning{Dresing et al.}

   \author{N. Dresing
          \inst{1}
                \and
          J. Gieseler\inst{1}
          \and
          M. Baumgartner-Steinleitner\inst{6}
          \and
          C. Briand\inst{10}
          \and
          K. Dissauer\inst{6}
          \and
          A. Fedeli \inst{1}
                    \and
          R. Jarolim\inst{6, 14}
                    \and
          J. T. Lang\inst{1}
          \and
          D. E. Morosan\inst{1,2}
          \and
          J. A. J. Mitchell\inst{3}
                    \and
          C. Palmroos\inst{1}
          \and
          J. Pomoell\inst{5}
                    \and
          J. Zhang\inst{10}
          \and
          O. Santala\inst{1}
          \and
          A. M. Veronig\inst{6}
          \and
          L. Vuorinen\inst{4}
          \and
          A. Warmuth\inst{3}
                O. Flor\inst{3,8}
          \and
          M. Flossie\inst{9}
                    \and
          E. K. J. Kilpua\inst{5}
                    \and         
          W. P\"otzi\inst{7}
          \and
          D. J. Price\inst{5}
          \and
          A. Gomes\inst{11,12}
          \and
          E. J. Hotti\inst{5}
          \and
          R. A. Hyndman\inst{13}
                    \and
          M. Liebel\inst{3,8} 
                    \and
          A. Razquin\inst{6}
                    \and
          S. Tan\inst{3,8}
                    \and
          R. Vainio\inst{1}
          }

   \institute{Department of Physics and Astronomy, University of Turku, Finland\\
              \email{\href{mailto:nina.dresing@utu.fi}{nina.dresing@utu.fi}}
        \and
             {Turku Collegium for Science, Medicine and Technology, University of Turku, 20014, Turku, Finland}
    \and
    {Leibniz Institute for Astrophysics Potsdam (AIP), An der Sternwarte 16, 14482 Potsdam, Germany}
    \and
    {Department of Physics and Astronomy, Queen Mary University of London, UK}
    \and
    {Department of Physics, University of Helsinki, Finland}
    \and 
    {University of Graz, Institute of Physics, Universitätsplatz 5, 8010 Graz, Austria} 
    \and
    {University of Graz, Kanzelh\"ohe Observatory for Solar and Environmental Research, Kanzelh\"ohe 19, 9521 Treffen, Austria} 
    \and 
    {Institut für Physik und Astronomie, Universität Potsdam, Karl-Liebknecht-Straße 24/25, 14476 Potsdam, Germany}
    \and
    {Centre for mathematical Plasma-Astrophysics, Department of Mathematics, KU Leuven, Celestijnenlaan 200B, 3001 Leuven, Belgium} 
    \and{LIRA, Observatoire de Paris, Université PSL, Sorbonne Université, Université Paris Cité, CY Cergy Paris Université, CNRS,  92195 Meudon, France}
    \and{Laboratory for Instrumentation and Experimental Particle Physics (LIP), Lisboa, Portugal}
    \and{Instituto Superior Técnico, University of Lisbon, Lisbon, Portugal}
    \and{Jeremiah Horrocks Institute, University of Central Lancashire, Preston, United Kingdom}
    \and{High Altitude Observatory, NSF National Center for Atmospheric Research, USA}
    }


   \abstract{}{}{}{}{}        
 
  \abstract
   {Solar eruptions comprise of a multitude of phenomena such as flares, coronal mass ejections (CMEs), large-scale coronal waves, radio bursts, and energetic particles traveling through interplanetary space. These phenomena are observed with a variety of instrumentation, including remote sensing and in-situ detectors. Obtaining a global understanding of a solar eruption often requires the analysis of various of these different datasets, including a multitude of analysis and modeling tools and a wide range of expertise. Usually, such a comprehensive analysis can only be achieved by a skilled and broad team. 

    The Energetic Solar Eruptions: Data and Analysis Tools (SOLER) project aims at creating a comprehensive analysis platform for the study of solar eruptions that allows a single user to easily apply analysis methods addressing various counterparts of the solar event. 

    Therefore, each partner of the project developed Python-based software, including interfaces in the form of Jupyter Notebooks, which provides application examples and concise step-to-step documentation. 

    In this paper we introduce the comprehensive solar-eruption-analysis infrastructure developed within the SOLER project. We explain where to find the software, how to use it, and give dedicated use-case examples of how to employ selected tools in a combined manner. 

   }        


   \maketitle
   
\newpage
\section{Introduction}

The goal of this paper is to present the solar eruption analysis infrastructure developed within the {\it Energetic Solar Eruptions: Data and Analysis Tools (SOLER)} project\footnote{\label{soler-website}\url{https://soler-horizon.eu}} We introduce the project's solar event catalogues and new analysis tools and showcase how they can be combined to perform a comprehensive solar eruption analysis. The SOLER project is aimed at investigating the most energetic phenomena occurring at the Sun to gain better understanding of the interrelations of various eruption phenomena, their variability and energy partitioning. SOLER makes use of the expanded, unprecedented heliospheric spacecraft fleet consisting of Solar Orbiter \citep[][]{Muller2020}, Parker Solar Probe \citep[PSP;][]{Fox2016}, the ahead spacecraft of the Solar TErrestrial RElations Observatory \citep[STEREO~A;][]{Kaiser2008}, and near-Earth spacecraft such as the SOlar and Heliospheric Observatory \citep[SOHO;][]{Domingo1995}, Wind \citep[][]{Ogilvie1997}, and the Geostationary Operational Environmental Satellite \citep[GOES][]{}, and BepiColombo \citep[][]{Benkhoff2021} on its cruise to Mercury. Observations of this fleet are employed to investigate energetic solar eruptions of solar cycle 25 starting from three perspectives: fast ($>$1000 km/s) coronal mass ejections (CMEs), strong ($>$M5) X-ray flares, and large solar energetic particle (SEP) events (reaching proton energies of 25 MeV). Key parameters of these eruption phenomena have been compiled into three, interlinked event catalogues \citep{morosan_2026} providing an excellent resource for event selection.

 SOLER's open-source analysis tools, which primarily take the form of Jupyter Notebooks, can be freely accessed via the project community on GitHub or used directly on the dedicated project JupyterHub server. They cover all aspects of a solar event, from the flare at Sun all the way to the energetic particle event measured in interplanetary space. Specifically, the novel data analysis and visualization tools allow studying the X-ray flare, UV and EUV features such as large-scale coronal waves, flare ribbons and coronal dimming, the structure of the corona resulting from MHD modeling, radio bursts, magnetic connectivity from the source to the spacecraft, and in-situ observations, most importantly measurements of solar energetic particles. This newly formed comprehensive infrastructure is openly available to the scientific community facilitating hassle-free interdisciplinary studies of energetic solar eruptions.

In Sect. \ref{sec:catalogs} we introduce the three interlinked energetic events catalogues of the project and in Sect.~\ref{sec:tools} we present the overall SOLER tools infrastructure followed by detailed descriptions of the various tools and a short guide on where to access the tools. We then discuss a few specific analysis use cases in Sect.~\ref{sec:use_cases}, where we describe how results of the different SOLER tools can be combined to study specific aspects of a solar eruption. 

\begin{figure}[t!]
    \centering
    \includegraphics[width=0.95\linewidth]{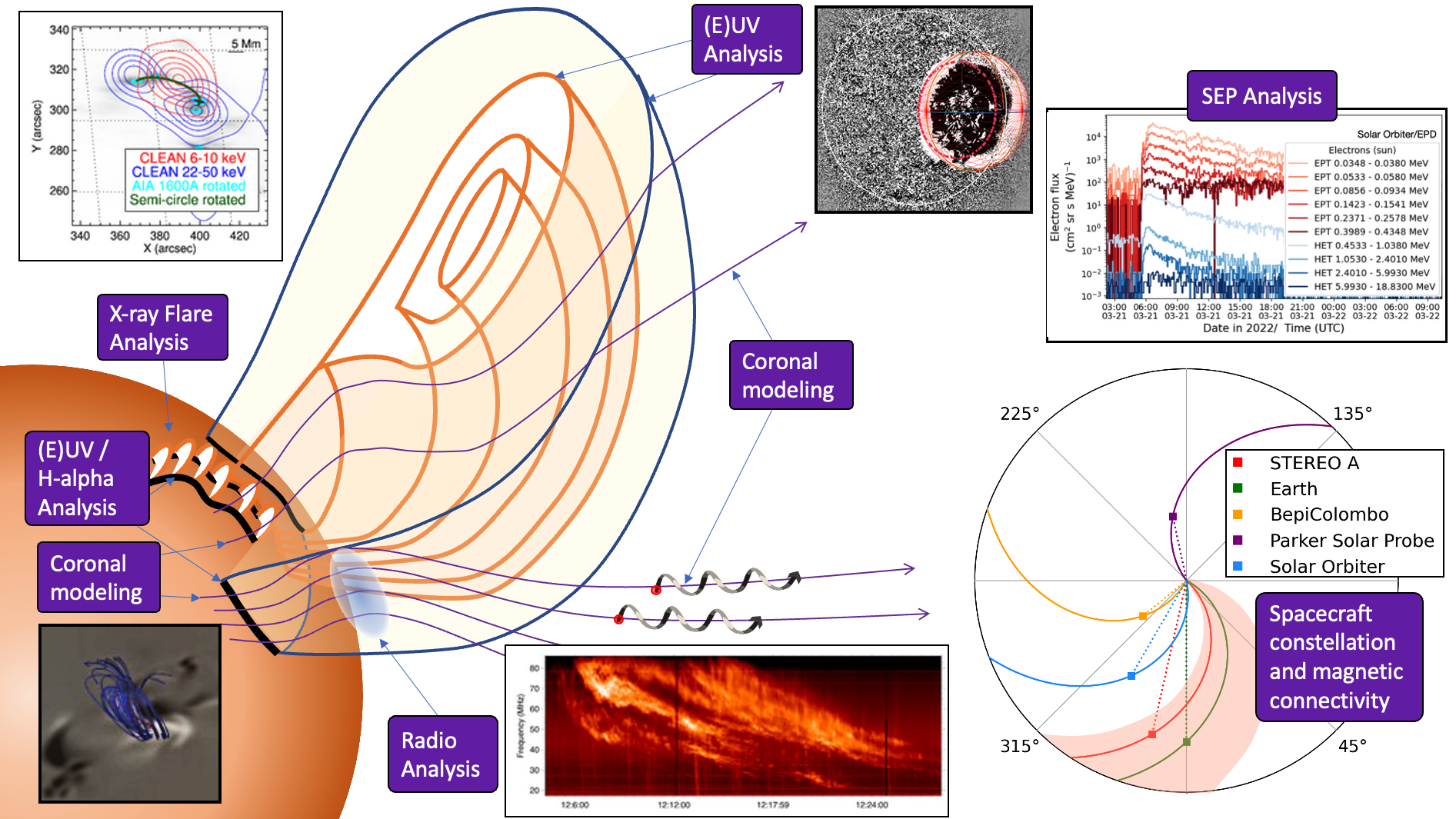}
    \caption{The SOLER analysis concept. Solar eruptions (flares, CMEs, SEP events) are  analyzed through a multi-messenger approach and by connecting the various aspects of the eruption with the set of analysis and visualization tools developed especially for integrated use. The violet boxes indicate the different methodology regimes of the project. They and appear again in Fig.~\ref{fig:workflow}, which provided an overview on the available analysis tools.}
    \label{fig:cartoon}
\end{figure}

\begin{figure}[t!]
    \centering
    \includegraphics[width=\linewidth]{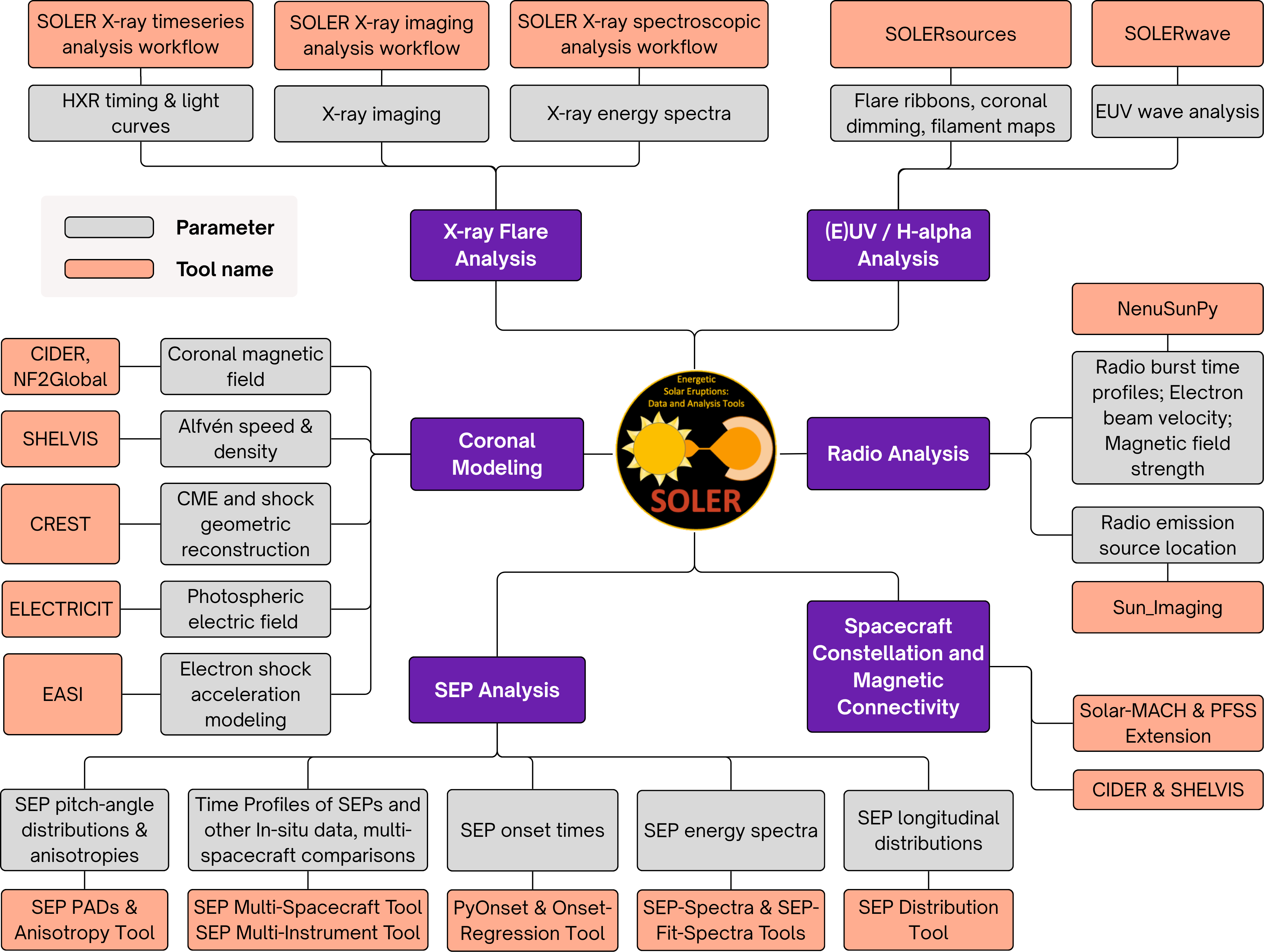}
    \caption{Flowchart illustrating the different analysis regimes (violet boxes) and related parameters (gray boxes) and tools (orange boxes) of the SOLER project allowing for a comprehensive solar eruption analysis from the Sun to the interplanetary spacecraft. The violet boxes are the same as in Fig.~\ref{fig:cartoon}.}
    \label{fig:workflow}
\end{figure}

\section{SOLER Catalogues and new Data Products}\label{sec:catalogs}


{Three new and interlinked catalogues have been released as part of the SOLER project \citep{morosan_2026} and made available through an interactive web-app\footnote{\url{https://soler-eu.streamlit.app}}. The catalogues consist of strong solar flares ($>$M5), fast ($>$1000 km/s) CMEs and SEP events where protons with energies $>$25 MeV are observed. The flares included are observed from the Earth’s orbit (GOES) and Solar Orbiter, the CMEs are observed at L1 (SOHO) and the SEP events are observed at five different locations: Solar Orbiter, PSP, STEREO~A, L1 (Wind and SOHO), and BepiColombo.} 

{For the strong flare selection, the catalogues contain information from solar event reports provided by the National Oceanic and Atmospheric Administration (NOAA) Space Weather Prediction Centre (SWPC). The flares with a class $>$M5 and their properties are included in the SOLER flare catalogue. In addition to the flare events observed from Earth, we also used the flare list from the Spectrometer/Telescope for Imaging X-rays \citep[STIX;][]{stix2020} onboard Solar Orbiter\footnote{\url{https://datacenter.stix.i4ds.net/view/flares/list}}. From this list, the catalogues also contains the flares observed by the STIX instrument that have an estimated flare class $>$M5, which is based on the GOES flare classification. The GOES and STIX flares are combined into a single comprehensive flare catalogue from two observers. In the case of fast CMEs, the SOLER catalogue uses the existing CME lists from the Large Angle and Spectroscopic Coronagraph \citep[LASCO;][]{br95} onboard SOHO  at Lagrangian point L1 with speeds $>$1000 km/s. For the SEP event selection, in situ observations from five spacecraft located at different observing points around the Sun were used to construct the SOLER SEP catalogue. The catalogue provides SEP characteristics for $>$25 MeV protons, 1 MeV electrons, and 100 keV electrons.}

{The three catalogues are linked in order to identify common energetic events and to provide key parameters of flares, CMEs and SEPs. At the time of writing, the catalogues contained 367 strong STIX and GOES flares combined and 231 fast CMEs from the start of Solar Cycle 25 until 31 December 2024, and 98 $>$25~MeV proton events until 31 December 2023. The parameters can be searched to identify, for example, useful case studies or combined across each or all catalogues to carry out statistical studies on energetic eruptions. The catalogues also contain information about the presence of metric type II radio bursts, which are observed by ground-based  stations (noted as ``GB'' in the catalog), associated with each flare, CME or SEP event. Type II radio bursts are important remote radio emissions as they indicate presence of a propagating coronal shock \citep[e.g.,][]{morosan2025}. These catalogues also provide the opportunity to study energetic events that do not always have the strongest flare, fastest CME or high-energy SEPs in common.}


\section{SOLER Tools Infrastructure}\label{sec:tools}
The cartoon shown in Fig.~\ref{fig:cartoon} displays the various solar eruption analysis regimes covered by the SOLER infrastructure. Starting close to the Sun, these include analysis methods for X-ray, (E)UV, and radio observations. Together with various coronal modeling tools they allow us to characterize the solar eruption close to the Sun and in the lower corona. The corresponding solar energetic particle event detected by interplanetary spacecraft can be studied using various SOLER SEP-and in-situ data tools. A further important component of the SOLER infrastructure is the study of the magnetic connectivity between the solar phenomena and the observing spacecraft. The violet boxes in Fig.~\ref{fig:cartoon} present where the six different analysis regimes of the SOLER project are at play. For each of these regimes SOLER provides a variety of analysis tools, which are illustrated in Fig.~\ref{fig:workflow} branching out from the same violet boxes marking the six different analysis regimes. The gray boxes in Fig.~\ref{fig:workflow} indicate the key parameters determined by each tool, whose names are presented in the orange boxes. A strong focus lies on studying the solar eruption at the Sun and lower corona (upper part of Fig.~\ref{fig:workflow}), which is covered by various observational and modeling tools. These include the \textit{X-ray Flare Analysis}, the \textit{(E)UV / H-alpha Analysis}, and the \textit{Radio Analyses} regimes, as well as \textit{Coronal Modelling} covering simulations of the coronal plasma and magnetic field, and an electron shock acceleration model.  
The bottom left of the figure represents the \textit{SEP Analysis} regime, which contains various in-situ analysis tools mainly focusing on energetic particle observations. To connect the solar eruption phenomena with the energetic particle observations, the \textit{Spacecraft Constellation and Magnetic Connectivity} regime covers suitable tools. However, this interconnection is often not a trivial task and requires a more in-depth analysis involving further comparison and interconnections of the results obtained with the separate tools presented in Fig.~\ref{fig:workflow}. In Sect.~\ref{sec:use_cases} we will discuss several use case examples where several of the tools are combined in a specific analysis. In the following sections we will describe all SOLER tools in detail organized along the six analysis regimes discussed above.

\subsection{Accessing SOLER tools}\label{sec:hub}

The primary source for the tools is the SOLER Zenodo community\footnote{\url{https://zenodo.org/communities/soler}}, which provides each tool with a persistent DOI for long-term identification and impact-tracking. Secondary sources include the project GitHub community\footnote{\url{https://github.com/soler-he}} and the project website\footref{soler-website}. All software is hosted using a git version control system and links are provided in the following sections and collated in Table~\ref{tab:code}. Each repository contains its own documentation and installation instructions.

\begin{table}
    \centering
    \caption{The SOLER tools and their respective sources.}
    \label{tab:code}
    \begin{tabular}{lll}
        \toprule
        \textbf{Tool} & \textbf{DOI} & \textbf{Code repository with instructions}\\
        \toprule
         CIDER &  & \href{https://github.com/jpomoell/cider}{github.com/jpomoell/cider}\\
         \midrule
         CREST &  & \href{https://github.com/jpomoell/crest}{github.com/jpomoell/crest}\\
         \midrule
         EASI & {\href{https://doi.org/10.5281/zenodo.21309037}{10.5281/zenodo.21309037}} & \href{https://github.com/SeveNyberg/easi}{github.com/SeveNyberg/easi}\\
         \midrule
         ELECTRICIT &  & \href{https://github.com/jpomoell/electricit}{github.com/jpomoell/electricit}\\
         \midrule
         \multirow{2}{*}{\shortstack[l]{NenuFAR SUN\\Imaging workflow}} & \multirow{2}{*}{\href{https://doi.org/10.5281/zenodo.18848880}{10.5281/zenodo.18848880}} & \multirow{2}{*}{\shortstack[l]{\href{https://github.com/JingeZhang94/NenuFAR_SUN_Imaging_workflow}{github.com/JingeZhang94/}\\\href{https://github.com/JingeZhang94/NenuFAR_SUN_Imaging_workflow}{NenuFAR\_SUN\_Imaging\_workflow}}}\\
         & & \\
         \midrule
         NenuSunPy &  & \href{https://gitlab.obspm.fr/soler/nenusunpy}{gitlab.obspm.fr/soler/nenusunpy} \\
         \midrule
         NF2 & \href{https://doi.org/10.5281/zenodo.10564989}{10.5281/zenodo.10564989}& \href{https://github.com/RobertJaro/NF2}{github.com/RobertJaro/NF2}\\
         \midrule
         SEP tools & \href{https://doi.org/10.5281/zenodo.15058293}{10.5281/zenodo.15058293} & \href{https://github.com/soler-he/sep_tools}{github.com/soler-he/sep\_tools}\\
         \midrule
         SHELVIS &  & \href{https://github.com/jpomoell/shelvis}{github.com/jpomoell/shelvis}\\
         \midrule
         SOLERsources & \href{https://doi.org/10.5281/zenodo.21297054}{10.5281/zenodo.21297054} & \href{https://github.com/soler-he/SOLERsources}{github.com/soler-he/SOLERsources}\\
         \midrule
         SOLERWave & \href{https://doi.org/10.5281/zenodo.20644556}{10.5281/zenodo.20644556} & \href{https://github.com/soler-he/SOLERwave_tool}{github.com/soler-he/SOLERwave\_tool}\\
         \midrule
         \multirow{2}{*}{\shortstack[l]{SOLER X-ray\\analysis workflows}} & \multirow{2}{*}{\href{https://doi.org/10.5281/zenodo.21336236}{10.5281/zenodo.21336236}} & \multirow{2}{*}{\href{https://github.com/soler-he/sunkit-spex_demo/}{github.com/soler-he/sunkit-spex\_demo}}\\
         & & \\
         \bottomrule
    \end{tabular}
\end{table}

To reduce the barrier to entry for the use and deployment of SOLER tools as much as possible, the project also provides a JupyterHub server\footnote{\url{https://soler-horizon.eu/hub}}. This server has a selection of the tools and their associated Python environments pre-installed to allow use by anyone with a web browser. The installations are automatically updated when changes to the tools are made, ensuring that the latest versions are always available to the hub users. The hub also includes a user guide for new users who might not be familiar with Jupyter Notebooks. Currently, the only requirement for access is to have a GitHub account for verification. Notably, the server is hosted in collaboration with the concluded EU Horizon 2020 project SERPENTINE\footnote{\label{SERPENTINE}\url{https://serpentine-h2020.eu}} (Solar EneRgetic ParticlE aNalysis plaTform for the INner hEliosphere). As a result, users are provided with the tools from both projects, a wide ranging suite for heliospheric analysis.



\subsection{X-ray analysis}

The X-ray workflows presented in this paper \citep{jake_mitchell_2026_21336237} focus on the analysis of X-ray data taken by STIX, however the methodologies used are, in principle, applicable instrument agnostic and so can be applied to any HXR data. STIX is a hard-X-ray (HXR) imaging spectrometer with an energy range of 4-150~keV, and therefore provides diagnostics of the hottest flare plasma as well as probing the flare accelerated electron population temporally, spectrally, and spatially. Our workflows are split into three distinct analysis regimes, timeseries, spectroscopy and imaging. 

\subsubsection{X-ray timeseries analysis workflow}

The energy released by magnetic reconnection during a solar flare accelerates electrons from the corona down towards the solar surface where they impact the denser layers of the chromosphere and emit non-thermal bremsstrahlung radiation which dominates the HXR signal. Therefore, the temporal evolution of the HXR lightcurves provides insights into the processes involved in electron acceleration and in magnetic reconnection itself. HXR lightcurves are highly variable and show characteristic pulses. They are frequently observed to display Quasi-Periodic Pulsations \citep[QPPs; see][for a review]{Zimovets2021}. The mechanisms responsible for QPPs are not yet well understood, but some possible explanations include direct modulation of emission caused by MHD waves, modulation of the efficiency of the energy release process and spontaneous quasi-periodic energy release.

The SOLER X-ray timeseries analysis workflow is built around three tools: wavelet analysis, Automated Flare Inference of Oscillations (AFINO), and a Gaussian decomposition tool. This workflow combines these tools allowing for STIX HXR lightcurves to be analysed and the results from each methodology compared. 

\begin{figure}
    \centering
    \includegraphics[width=\linewidth]{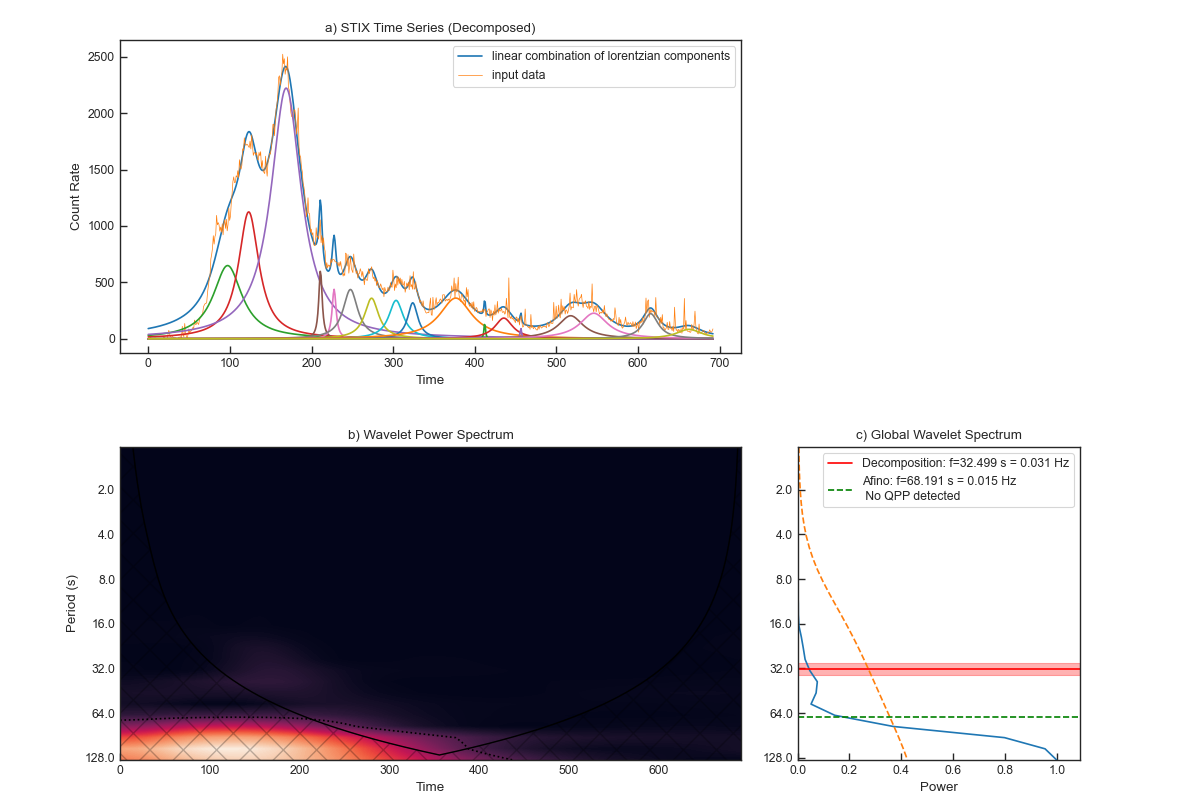}
    \caption{Examples of HXR time series analysis for an X3 flare on 5 Aug 2024. Top: The HXR lightcurve for this event decomposed into a combination of lorenztian profiles. Bottom left: The wavelet power spectrum for this lightcurve, with the area of significance outside of the hatched shading. Bottom right: The same wavelet analysis shown as period against power with all three methods represented for comparison, wavelet, decomposition and AFINO.}
    \label{fig:xray_timeseries}
\end{figure}

The decomposition methodology is based around the Gaussian decomposition tool\footnote{\url{https://github.com/hannahc243/Gaussian_Decomp}} described in \citet{Collier_2023}. The pipeline first smooths the observed signal using Gaussian Process regression before detecting peaks. Once candidate peaks have been identified, different line profiles (Gaussian, Lorentzian, and asymmetric Gaussian) are fitted to the bursts and the best fit identified. From this, the pipeline can provide estimates of burst timings, amplitudes, and periodicities, which can subsequently be interpreted in the context of possible QPP behaviour.

AFINO\footnote{\url{https://github.com/aringlis/afino_release_version}} analyses the Fourier power spectrum of HXR flare lightcurves. The code fits several power-law background models, some of which include a Gaussian component in the frequency spectrum that may indicate periodicities. Model selection is performed using the Bayesian Information Criterion (BIC), which balances fit quality against model complexity. A strong preference for models containing a Gaussian enhancement is interpreted as evidence for a QPP detection and provides an estimate of the dominant period.

In contrast to the global Fourier-based approach of AFINO, the wavelet method\footnote{\url{https://github.com/ct6502/wavelets}} performs a time--frequency decomposition of the signal using a localised Morlet wavelet. This method identifies oscillatory behaviour at different timescales and at different times throughout the observation by continuously stretching and translating the wavelet across the time series. The resulting wavelet power spectrum reveals both the dominant periods and how they evolve over time. Statistical significance is assessed relative to a red-noise background spectrum, whilst the cone of influence defines the regions where edge effects make the results unreliable. The wavelet analysis therefore complements AFINO by providing localised information on non-stationary QPP signatures.


Together, these three approaches provide complementary diagnostics: the decomposition pipeline identifies bursts in the time domain, AFINO searches for statistically significant global periodicities in the frequency domain, and the wavelet analysis investigates how potential periodicities evolve temporally throughout the flare. An example of the analysis from the workflow and a comparison of these three methodologies is displayed in Figure~\ref{fig:xray_timeseries}.


\subsubsection{X-ray spectral analysis workflow}

The SOLER X-ray spectral analysis workflow is primarily constructed around sunkit-spex 
\citep{sunkit_spex_2026} and STIXpy \citep{stixpy_2026}. Sunkit-spex is a package for the high-energy spectroscopic analysis of solar flares, whilst STIXpy is a package for the acquisition, calibration and imaging analysis of STIX data, both packages are Python based, open source and open development. 

The physical models used to characterise the different elements of flare emission yield photon fluxes, whereas spectra taken by high-energy instruments, such as STIX, are output in instrument specific count space. Therefore, spectroscopic analysis of high-energy data requires the use of instrument specific non-diagonal response matrices to convert physical models from photon spectra to instrument modeled count spectra that can then, through forward fitting and other methods, be fit to observed data. This requires the use of a dedicated software package such as sunkit-spex, which contains solar specific models, an instrument agnostic data container, and provides the ability to carry out time resolved and simultaneous fitting. 

Using the newly developed version of sunkit-spex alongside the recently implemented calibration and data reduction features in STIXpy, it is possible for the first time, as presented in this SOLER workflow, to fully acquire, calibrate and analyse STIX data in Python.

Firstly STIXpy is used to acquire the data from an online database for a specific observation time or observation ID. The downloaded science and background data can then be calibrated and given a specified time interval reduced into a 1D spectrum to be housed by the instrument agnostic sunkit-spex spectrum container. Along with the data itself, this spectrum object also contains all of the auxiliary information necessary to analyse X-ray solar flare data, including the livetimes, flare angle, and event specific non-diagonal spectral response matrix or SRM. The STIX SRM is specific to each event as it takes into account the event location in order to accurately characterise the grid transmission. This spectrum object can then be passed off to the sunkit-spex fitter object along with the desired model combination to be fit to the data. 

\begin{figure}
    \centering
    \includegraphics[width=\linewidth]{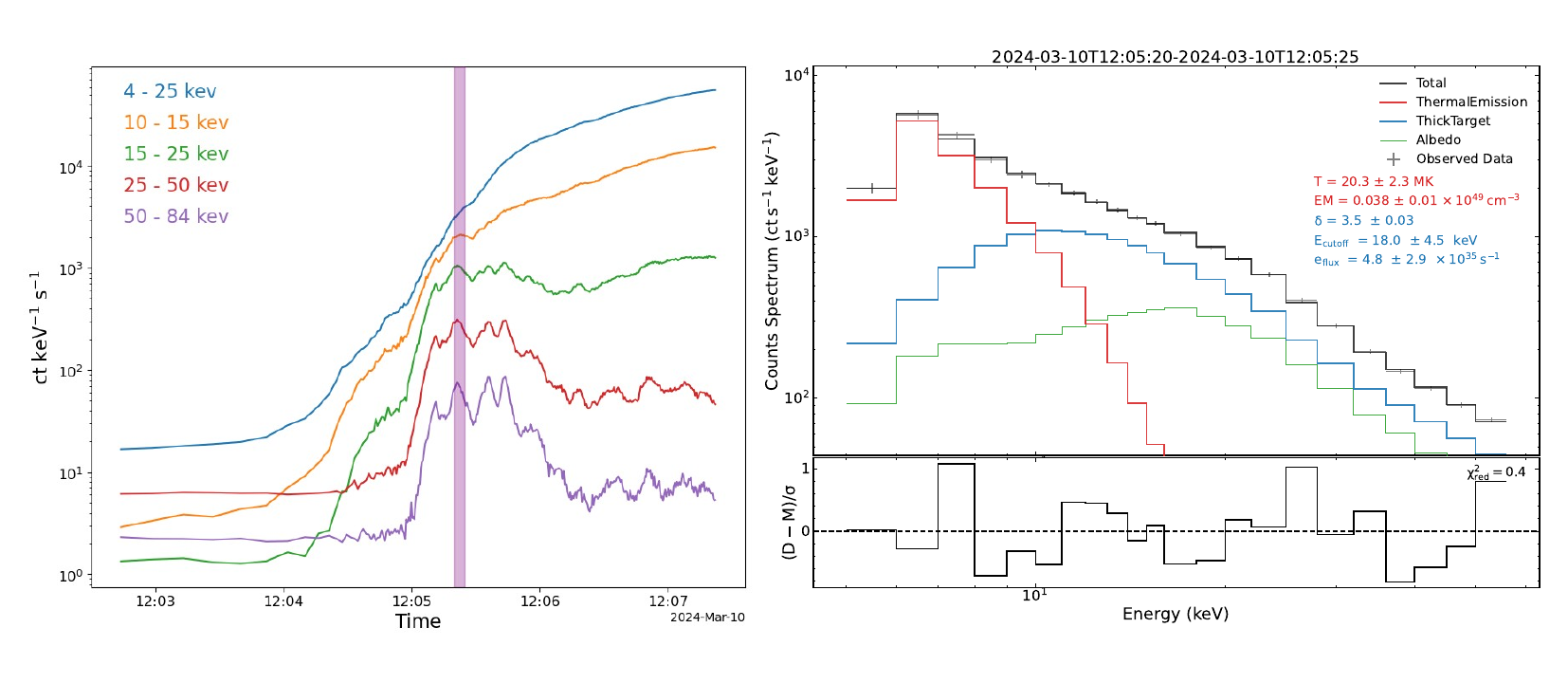}
    \caption{Left: Light-curves in  five different energy bands for the M7 flare of 10 March 2024. The integration time for the spectral analysis shown in the right-hand panel is highlighted in purple.  Right: Top panel: Observed count rates (grey points), fitted with a composite model (black line) comprised of a thermal model (red) a non-thermal model (blue) and an albedo model (green). The best fit parameters and their uncertainties are displayed in colours matching the respective emission component. Bottom Panel: The normalised residuals for each energy bin, along with the overall fit statistic.}
    \label{fig:xray_spec}
\end{figure}

The X-ray spectral analysis of solar flares, particularly when considering the impulsive peak of the flare which contains significant HXR contributions, typically requires the use of three main emission components, a thermal emission, a non-thermal thick target and an albedo component. Figure~\ref{fig:xray_spec} shows these three model components fit to the impulsive peak of an M7 event on 10 March 2024. Ultimately this fitting procedure allows the physical parameters which describe the thermal and non-thermal emission resultant from the flaring process to be constrained. This process can then be iterated to produce time series of these physical parameters giving insights into the evolution of the flare. 

\subsubsection{X-ray imaging analysis workflow}

The SOLER X-ray imaging workflow, built around STIXpy and XRAYVISION \citep{xrayvision} uses Fourier based techniques to construct images of the X-ray sources observed by STIX \citep{Massa2023}. The X-ray sources in the soft X-ray (SXR) regime and the HXR regime show distinct morphologies and evolve differently across the duration of the flare. The SXR sources typically form a single extended structure and characterise the thermal emission from the flare loop and so form gradually during the flare. Whereas the HXR sources are generally split into two or more distinct footpoints which represent the sites at which electrons accelerated down either side of the flare loops impact the dense layers of the chromosphere. These footpoints are most prominent during the initial impulsive peak of the flare, and often fade as the SXR source becomes more dominant. 

\begin{figure}
    \centering
    \includegraphics[width=\linewidth]{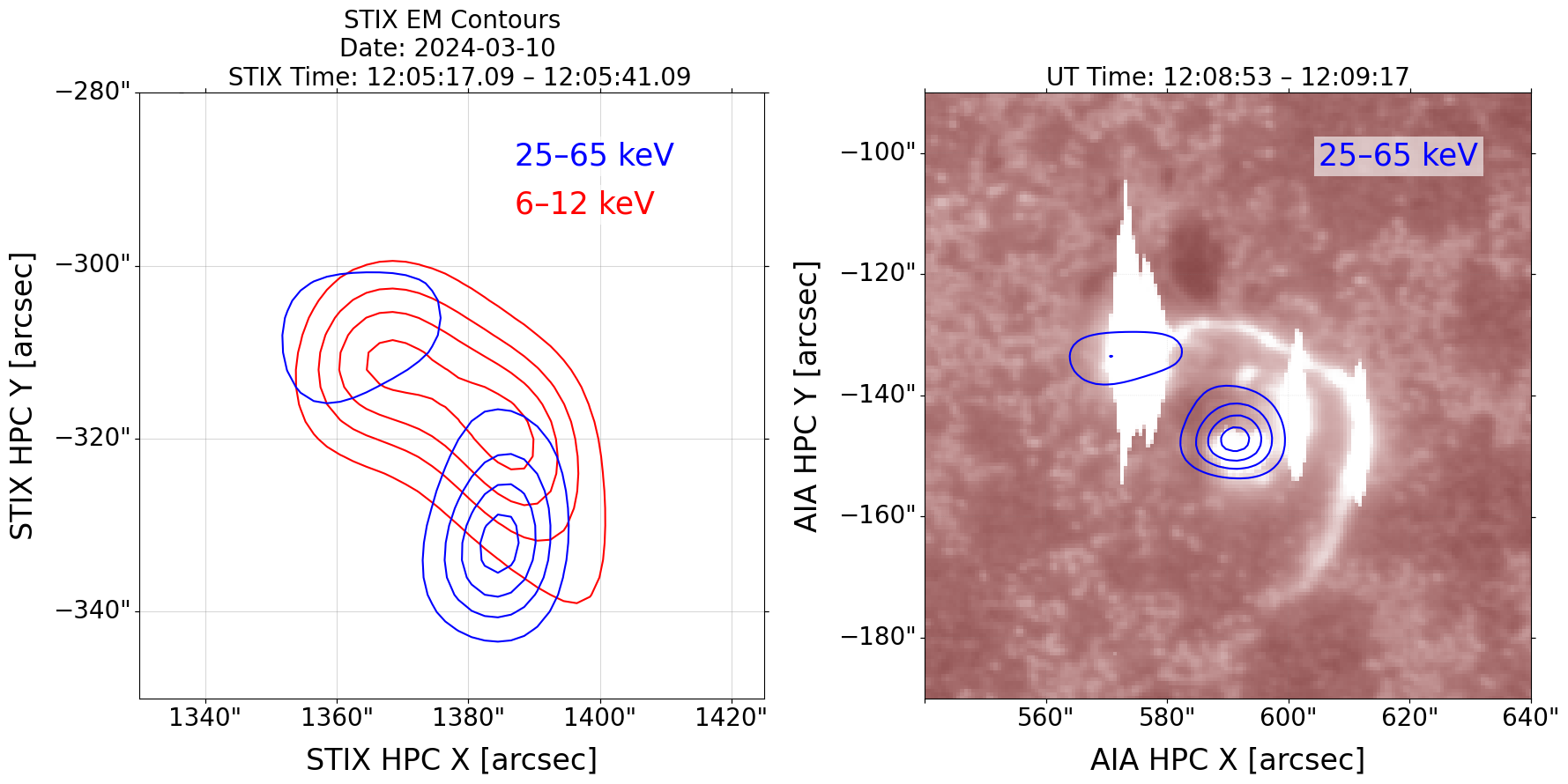}
    \caption{Imaging of the M7 flare on 10 March 2024. Left: STIX contours generated using the Expectation Maximisation (EM) algorithm in STIXpy. Two energy ranges are displayed, 6-12~keV in red, and 25-65~keV in blue. Right: The STIX 25-65~keV contours shown in the left-hand panel reprojected to the AIA 1700~\AA{} image for the same integration window.}
    \label{fig:xray_image}
\end{figure}

To create such images, a STIX compressed pixel data science and background file are used, from which a specific time integration window can be selected along with a specified energy range. The STIXpy workflow can then be used to locate the flare on the solar disc and then create a higher resolution map of the image. These maps are created using three different imaging algorithms, clean \citep{Hoegbom1974}, maximum entropy \citep[MEM GE;][]{Massa2020}, and expectation maximisation \citep[EM;][]{Massa2019}. The results from these algorithms can then be compared. The notebook also allows for the acquisition and creation of a Sunpy map for EUI and AIA data. These Sunpy maps can then be reprojected and overplotted. An example of EM maps of the M7 flare on 10 March 2024 is shown in Figure~\ref{fig:xray_image}. The panel on the left is from the non-thermal peak of the flare and shows the classic morphology, with two non-thermal footpoints seen at 25-65~keV (blue contours) and the SXR looptop source at 6-12~keV (red contours). In the right panel, the nonthermal sources have been reprojected and overplotted onto a co-temporal AIA 1700~\AA{} image.

\subsection{(E)UV analysis}

 \subsubsection{SOLERsources: flare, dimming and filament analysis}

One of the SOLER goals is to develop methodologies to create time series of solar images containing masks of three different features: filaments, flare ribbons and coronal dimmings, in order to facilitate the SEP analysis regarding the solar source regions and associated eruptive events. 

\textbf{Coronal dimming detection.}
Coronal dimmings are sudden decreases in the EUV and SXR emission of the Sun that occur during the early evolution of a CME. They are caused by the expansion-related density decrease and ejection of coronal plasma during a CME \cite[see the recent review by][]{Veronig2025}. The coronal dimming detection and analysis code is based on the detection algorithm and analysis developed in \cite{Dissauer2018a}. To identify coronal dimming regions and to follow their evolution, we apply a thresholding technique on logarithmic base-ratio EUV images from the Atmospheric Imaging Assembly \cite[AIA;][]{Lemen2012} onboard the Solar Dynamics Observatory \cite[SDO;][]{Pesnell2012} using its 211~{\AA} filter. The data is rebinned to $2048 \times 2048$ pixels under the condition of flux conservation and SunPy software \citep{sunpy_community2020} is used to perform image calibration. We check for constant exposure time and exclude AIA images where the automatic exposure control algorithm was triggered. In addition, we correct for differential rotation by rotating each image frame to a common reference frame at the beginning of the time series, chosen at 30~min before the start of the associated GOES flare. This reference frame also acts as base image for the detection and represents the median over the first 10 images within the time series in each pixel. All pixels whose logarithmic (log10) ratio intensity decreased below $-0.19$ corresponding to a change of about 35\% in linear space, are identified as dimming pixels. To reduce noise and the number of misidentified pixels, morphological operators are used to smooth the extracted regions, i.e., small features are removed, while small gaps are filled.

\begin{figure}
    \centering
    \includegraphics[width=0.8\linewidth]{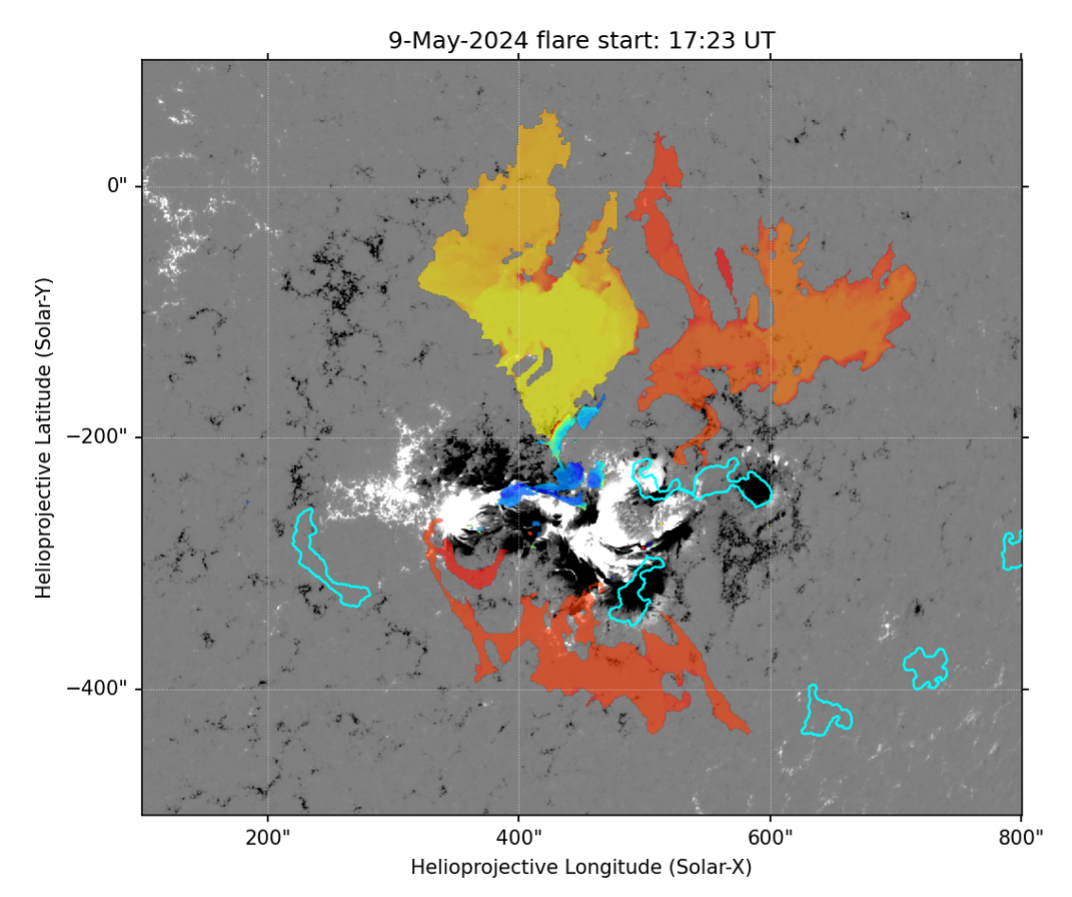}
    \caption{Combined flare ribbon (in rainbow color), coronal dimming (dark orange to orange) and filament masks (cyan contours) on top of representative SDO/HMI line-of-sight magnetogram for the flare/CME event on May 9, 2024.
    }
    \label{fig:combined-mask}
\end{figure}

Dimming regions are, in general, complex, and different parts may grow and recover on different timescales. From instantaneous dimming masks $D_k(p_i,t_k)$, representing all dimming pixels $p_i$ detected at a specific time step $t_k$, we calculate cumulative dimming pixel masks $C_m(p_i,t_m)$ by combining all dimming pixels identified up to the time $t_m$. In this way, we can capture the full extent of the total dimming region over time (for an example of a cumulative dimming mask see Fig.~\ref{fig:combined-mask}). 
The original algorithm used in \cite{Dissauer2018a} was developed in IDL, and has been translated to Python (open source) for the purpose of the SOLER project.

\begin{figure}
    \centering
    \includegraphics[width=0.99\linewidth]{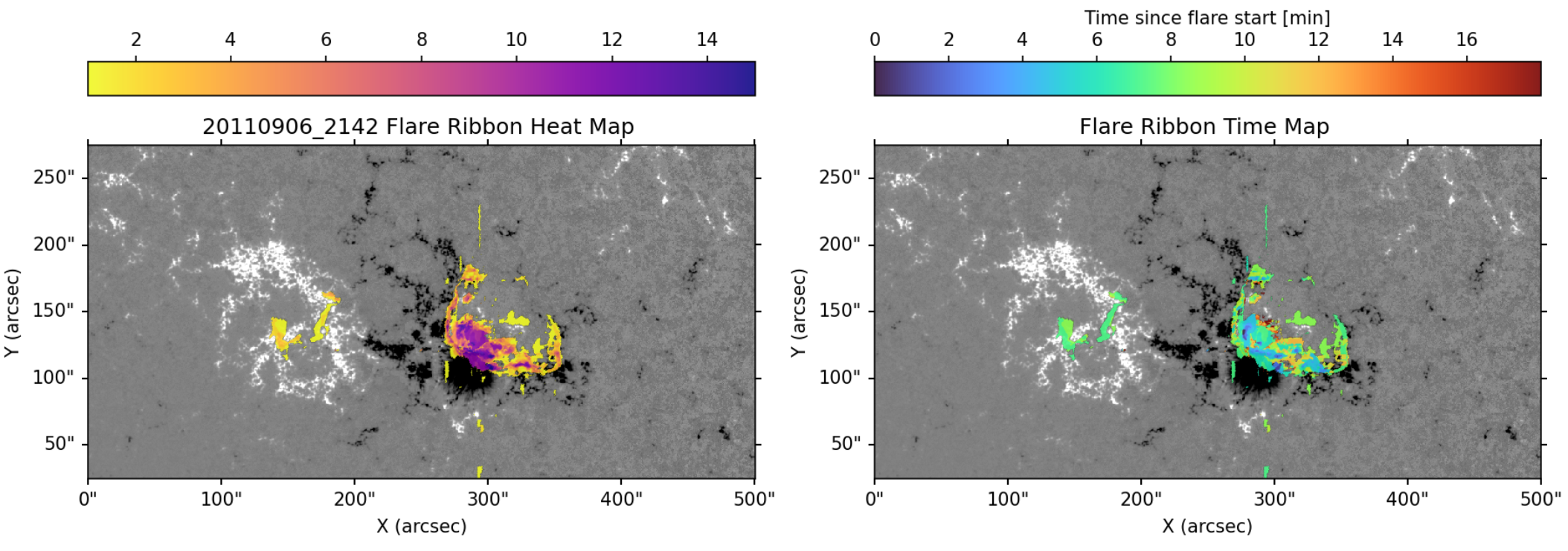}
    \caption{Flare ribbon heat map (left) and time map (right) on top of the radial component of a representative SDO/HMI vector magnetogram for the X2.1 flare on September 6, 2011. Colors in the heat map indicate how often a pixel was flagged as a flare pixel during the flare duration; colors in the time map indicate when each flare pixel was detected for the first time (in minutes since the start of the flare).}
    \label{fig:flare_heatmap}
\end{figure}

\textbf{Flare ribbon detection.}
For automatic detection of flare ribbons in SDO/AIA 1600~{\AA} UV filtergrams, we developed a Python code following the approach by \cite{Kazachenko2017}. As an initial step, SDO/AIA 1600~{\AA} images are corrected for image saturation due to CCD blooming. This is done by first selecting pixels above the saturation level, $I_{\rm sat}=4000$~counts s$^{-1}$, and the pixels surrounding them within 3 to 30~pixels in the $x$- and $y$-directions. Each saturated pixel intensity is then replaced with the value linearly interpolated in time between the individual pixel’s previous and subsequent unsaturated values. We additionally remove artifacts around these initially selected saturated pixels (independent of a threshold intensity) by using the median absolute deviation estimated from pre-flare images \citep{Maybhate2008}. 

In the detection algorithm, for each image $k$ at time step $t_k$, flare pixels are identified as pixels $p$ with intensity values higher than a predefined cutoff intensity $I_c$, which is a multiple $M$ of the median intensity of $k$ over a selected field-of-view around the flare location. $M \approx 8$ has been empirically found as lower limit for the detection of flare ribbons and is used here. 
Similar to coronal dimmings, we provide instantaneous and cumulative flare pixel masks to characterize the flare evolution. In addition, we create a heat map  and a time map   to study flare ribbon activity during the flare duration as defined by the start and end time of the corresponding GOES X-ray flare.
The heat map is based on a cumulative flaring mask, i.e. includes information over the entire flare duration and resolves hot spot areas of enhanced flaring activity. Flaring pixels that occur multiple times at the same location or are longer active have a higher value in this mask. The time map is also a cumulative map. In this case, each flaring pixel contains information on its first detection.
Figure~\ref{fig:flare_heatmap} shows examples of a flare ribbon heat map and time map for the X2.1 flare on September 6, 2011.

\begin{figure}
    \centering
    \includegraphics[width=0.85\linewidth]{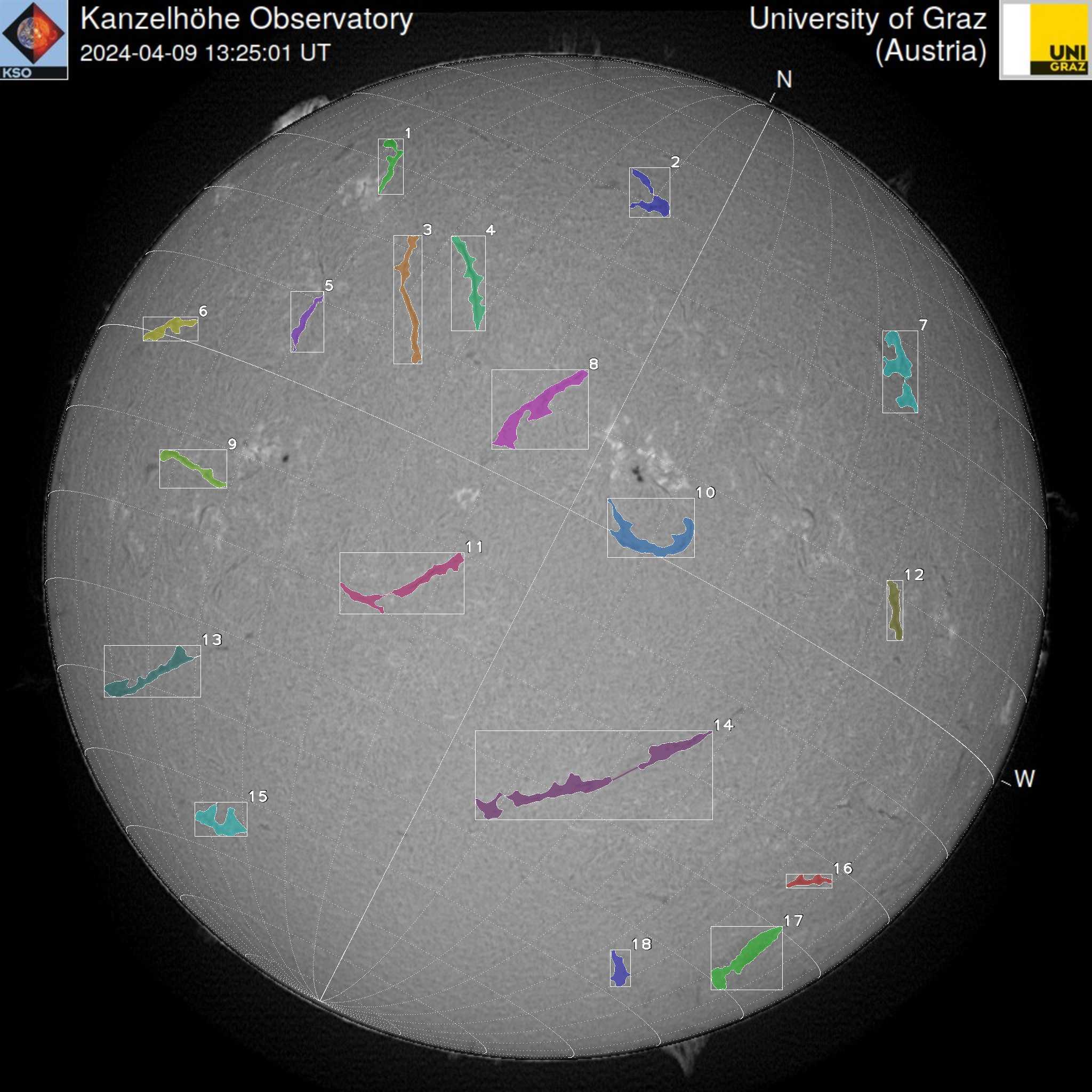}
    \caption{Example of a solar full-disk H$\alpha$ image from Kanzelh\"ohe Observatory, together with the identified filaments (in colour). Parameters of the filaments can be found at \url{https://cesar.kso.ac.at/filament/filament_data.php?date=20240409}
    }
    \label{fig:filament}
\end{figure}

\textbf{Filament detection.}
For the filament masks, we make use of the regularly provided data product from Kanzelh\"ohe Observatory of the University of Graz \citep{Poetzi2021}. Filaments are automatically detected on real-time H$\alpha$ filtergrams recorded at Kanzelh\"ohe Observatory using the detection algorithm described in \citet{Poetzi2015}. For each filament, the length, area, location and orientation of the filament axis is derived, and the segmentation masks are provided. 
An example H$\alpha$ image along with the detected filament masks is shown in Figure \ref{fig:filament}. 
In the framework of SOLER, we developed a new data product at Kanzelh\"ohe Observatory to provide the filament masks (\url{https://cesar.kso.ac.at/filament/filaments.php}) as FITS files, to easily combine them with other observations and data products such as the flare and coronal dimming masks. 

\textbf{Combined flare, dimming, filament masks.}
To facilitate in-depth SEP analysis regarding the solar source regions and associated eruptive events, we combine the derived masks of flare ribbons, coronal dimmings and filaments into one joint visual representation. Flare ribbons are in general detected over the flare duration, as defined by the GOES soft X-ray flare catalogue, coronal dimmings during a 1.5-hour time period, starting 30~min before the corresponding flare onset to cover most of the impulsive phase of the dimming, as found statistically by \cite{Dissauer2018b}. Since filaments are in general considered to be long-lived features, we select only a single time step to outline the filaments, choosing the observation from Kanzelh\"ohe Observatory that is closest to the flare onset.

Figure \ref{fig:combined-mask} shows an example of this combined visual representation. In order to better see the evolution of the flare and coronal dimming, we do not show the full solar disk (for which the analysis is performed) but a smaller field-of-view zoomed into the active region.  Filaments are indicated as cyan contours, while the time evolution of flare ribbons and coronal dimmings is shown in rainbow color (blue representing earlier, red representing later times) and different shades of orange (lighter orange representing earlier, darker orange representing later times), respectively. 

\emph{SOLERsources} is written in Python 3.13 using SunPy 7.0.3 \citep{sunpy_community2020} and Astropy 7.0.0 \citep{astropy:2013, astropy:2018, astropy:2022}. The code suite is available to the public via GitHub\footnote{\url{https://github.com/soler-he/SOLERsources}} \citep{SOLERsources_2026} under a BSD 3-Clause License. The code suite also contains an educational Jupyter Notebook to demonstrate the application to a flare/CME event to showcase the detection and segmentation of coronal dimmings, flare ribbons and filaments.

\subsubsection{Large-scale coronal wave analysis}
\begin{figure}
    \centering
    \includegraphics[clip, trim=1.75cm 2.75cm 2.25cm 1cm,width=0.9\linewidth]{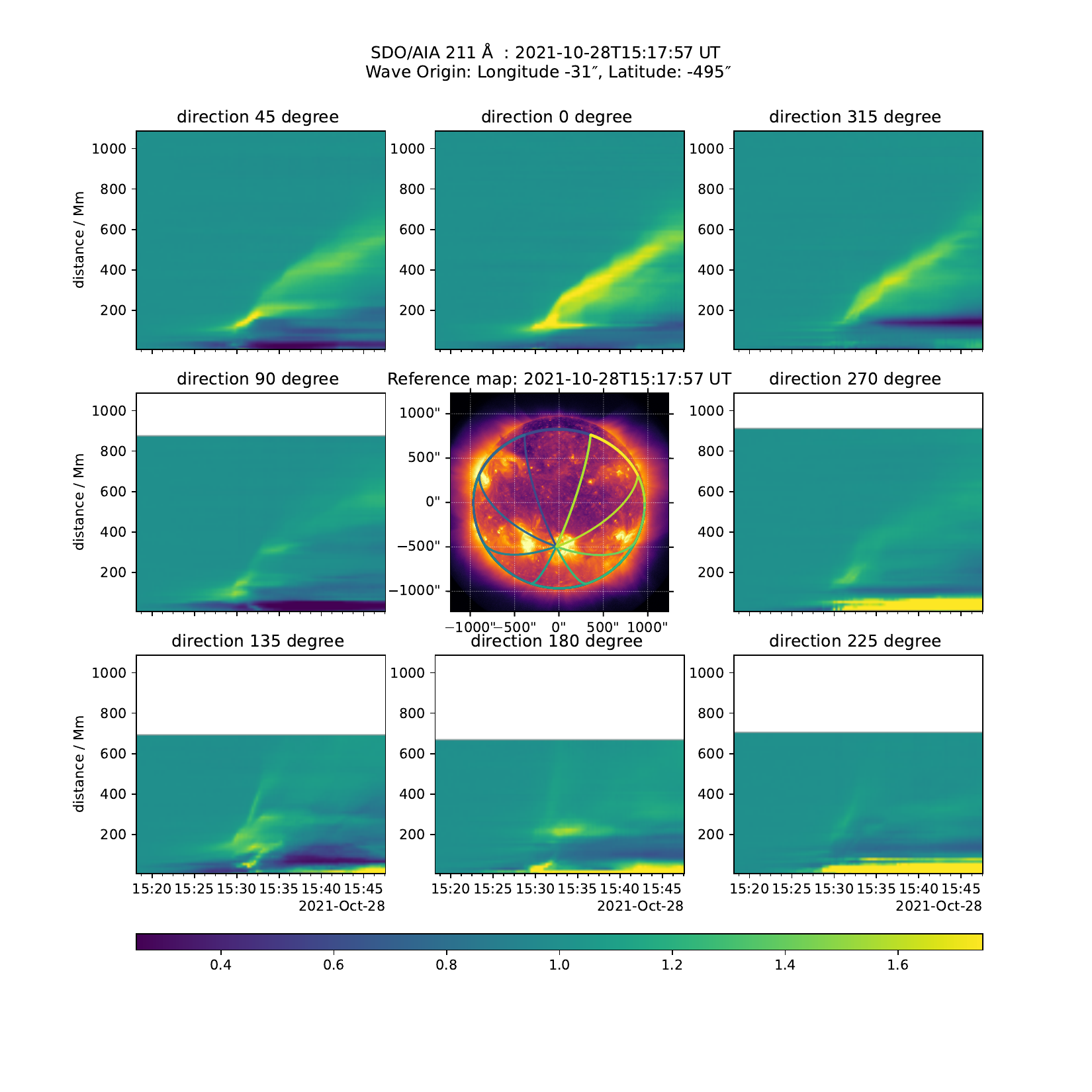}
    \caption{Overview plot for the coronal wave on 28th October 2021. The plot consists of J-maps showing the wave evolution in eight sectors, each of a width of 45° (the sector boundaries are illustrated in the base image shown in the middle panel). The J-maps show the intensity derived from SDO/AIA 211 {\AA} base ratio images. The color bar below quantifies the corresponding intensity values, which we restricted in the visualization to the range [0.3,1.7].}
    \label{fig:wave_octant}
\end{figure}

\begin{figure}
    \centering
    \includegraphics[clip, trim=1.25cm 1.5cm 1.75cm 1cm,width=0.9\linewidth]{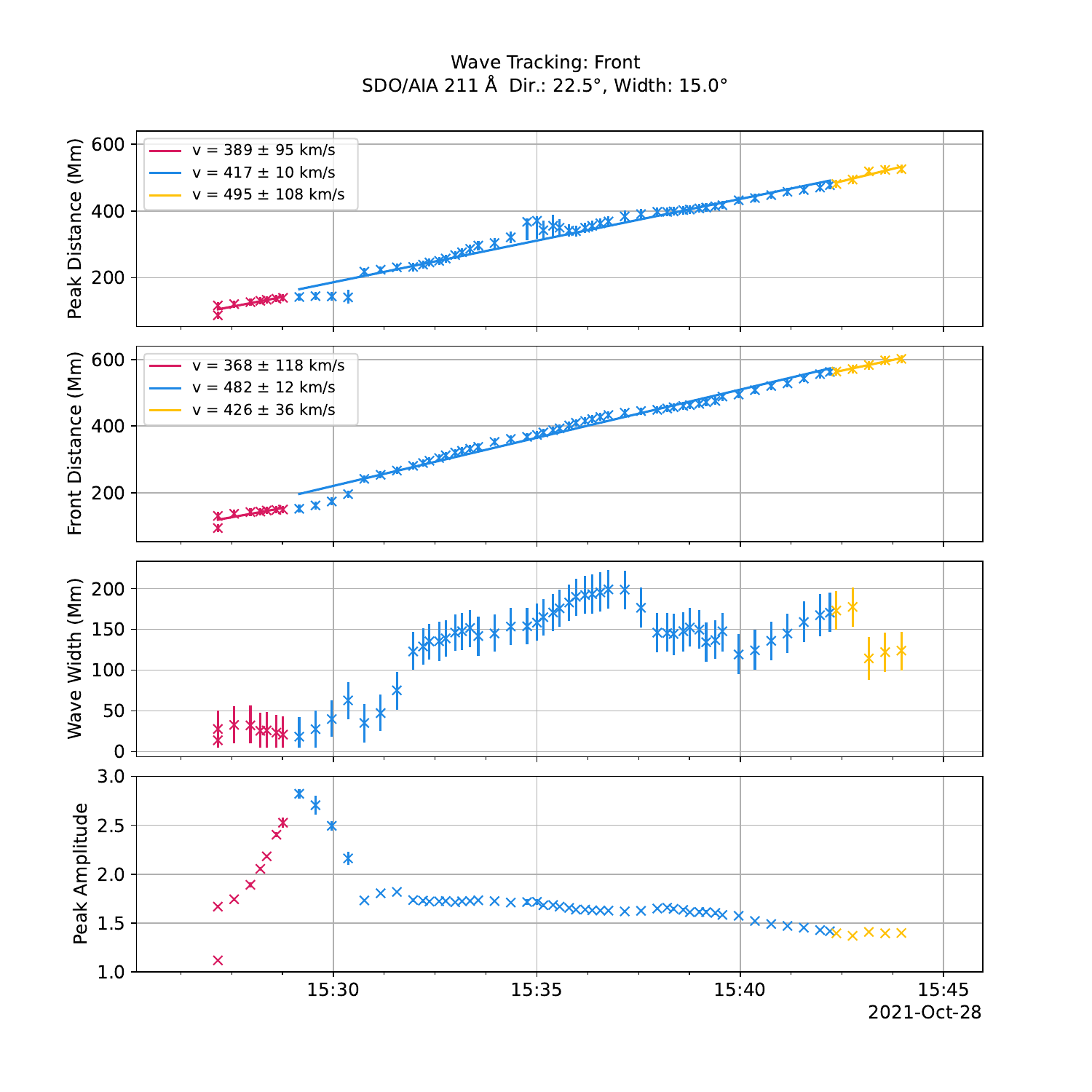}
    \caption{Kinematics plot for the large-scale coronal wave observed on the 28th October 2021 as derived by the SOLERwave tool. From top to bottom: distance of the peak of the wave, the front of the wave, the wave width, and wave amplitude calculated from SDO/AIA $211$~{\AA} images into the Northern direction (see Fig.~\ref{fig:wave_octant}).}
    \label{fig:wave_kinematics}
\end{figure}




For the analysis of large-scale coronal waves (or EUV waves) and Moreton waves, the SOLERwave tool \citep{Baumgartner_2026,SOLERwave_2026} was developed. It provides algorithms for automatically detecting large-scale waves in sequences of EUV or H$\alpha$ images, and for determining their kinematics, speeds and amplitude evolution. The SOLERwave tool is based on the calculation of perturbation profiles \citep[e.g.,][]{warmuth_evolution_2001,podladchikova_automated_2005, muhr_analysis_2011},
which utilize the increase in intensity due to the passage of a wavefront that compresses the plasma at the base of the corona and chromosphere. The perturbation profiles are derived along great arcs on the spherical solar surface, which are centered at a presumed wave origin. Large-scale waves manifest in these perturbation profiles as local enhancements that are propagating in time. To enhance the signature of these transient features, in the SOLERwave tool the perturbation profiles are derived from base ratio images, where each image of the sequence is divided by a pre-event image. 

The main workflow of the SOLERwave tool is split into two steps: file loading and data preprocessing, and the wave analysis itself. The first step is handled in the \textit{Load\_new\_event} Jupyter Notebook that downloads files and preprocesses them after the user specified the start and end time of the event under study and the instrument to be used.
The start time also marks the reference image used for the base ratio images and is typically chosen between $5$ to $10$~min prior to the flare onset.  The preprocessing includes standard image correction related to the instrument used (e.g., SDO/AIA, STEREO/EUVI), the correction of the images for solar differential rotation, binning of the images and calculation of the base ratio images. The \textit{Load\_new\_event} Jupyter Notebook also includes an example code on how to preprocess existing data, should it have been downloaded for a previous analysis. 

The second step is the wave analysis itself, using the SOLERwave tool as demonstrated in the \textit{SOLERwave\_Event\_Study} Jupyter Notebook. Besides the preprocessed data, the tool requires as user input the presumed wave origin, usually set to the position of the associated flare. Following the function calls as described in the Jupyter Notebook, the first visual output of the analysis is the so-called octant plot; an example is shown in Figure~\ref{fig:wave_octant}. The octant plot consists of eight J-maps derived from perturbation profiles in different directions. J-maps are plots with time on the $x$-axis, distance on the $y$-axis and the color encoding intensity (in our case, from base ratio images). Each J-map has been created from perturbation profiles calculated along sectors with a width of $45^\circ$ and a longitudinal resolution of $1$°. In these plots, the propagating wave appears as an oblique line of enhanced emission. Using this information, the user chooses a sector for further investigation and runs the remaining part of the Jupyter Notebook. This prompts a peak finding algorithm to automatically identify the peaks in the perturbation profiles created in the sector of investigation and a wave tracer to infer motion. If peaks in consecutive observations have continuously propagated within the limits of an upper and lower speed limit (by default between $100$~km~s$^{-1}$ and $2000$~km~s$^{-1}$, can be adapted by the user), they are connected and identified as a wave. 

Figure~\ref{fig:wave_kinematics} shows a standard output plot from the SOLERwave tool for the resulting wave kinematics. The figure shows from top to bottom the distance of the peak of the wave (from its origin), the distance of the wave front as well as the  width and peak amplitude of the wave as function of time. For a typical observation of a (shocked) wave, the amplitude first increases as the wave gets more and more shocked, and decreases in later phases (along with a broadening of the width) as the wave expands and potentially also energy is dissipated \citep{mann_simple_1995,mann_propergation_2023}. Waves detected are color coded and are fitted with a linear function. The resulting slope gives the mean propagation speed, which is annotated in the legend of the kinematics plot along with its uncertainty. Besides the two visual outputs shown in Figs.~\ref{fig:wave_octant} and \ref{fig:wave_kinematics}, the SOLERwave tool creates output plots with multiple panels showing the evolution of the perturbation profiles as well as a movie showing both the perturbation profiles in the selected sector and the corresponding base ratio images overplotted with the sector and the determined wave front and peak positions. 



\subsection{Radio burst analysis}
Solar eruptions release energetic particles that propagate through the corona and the interplanetary medium. Various mechanisms produce electromagnetic waves from this magnetized plasma. For instance, electron beams act as a source of free energy that destabilizes the space plasma, initially generating electrostatic waves that are subsequently converted into electromagnetic waves. Several conversion mechanisms produce radio emissions in the decameter-to-kilometer wavelength range. At lower altitudes in the solar corona, the presence of a stronger magnetic field more actively controls the radio emission processes. Once emitted, photons travel in a non-uniform plasma and are submitted to absorption and scattering. Magnetic field strength, beam velocity and density, coronal temperature, density fluctuations, and anisotropies are among the parameters that govern radio emissions. 
Several radio telescopes, spread in USA, Australia or the Netherlands, operate in the range 10 - 300~MHz. The last generation,  {\it New Extension in Nançay Upgrading LOFAR (NenuFAR)}, is implemented at the Radio Astronomy Observatory of Nan\c{c}ay in France \citep{Zarka15}. It provides observations at the highest sensitivity, spectral and temporal resolution. While NenuFAR operates in dedicated observational campaigns, the Nançay Decameter Array \citep[NDA;][]{Lecacheux2000,Lamy2018} conducts daily observations of the Sun. Within SOLER a Python package and visualization tool have been specifically developed for previsualization and processing of NenuFAR and NDA observations.
\subsubsection{Previsualization}
\begin{figure}
    \centering
    \includegraphics[width=0.9\linewidth]{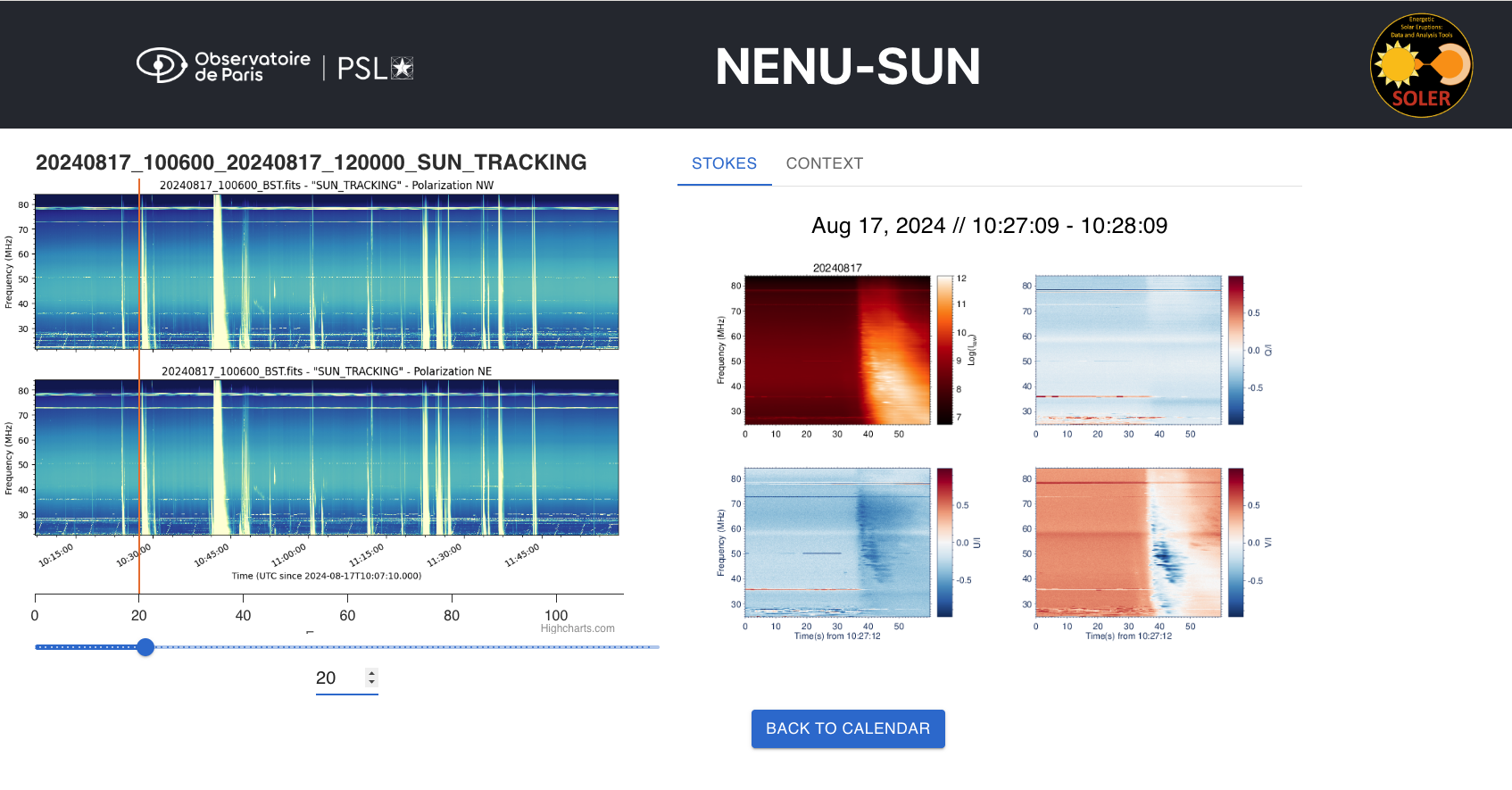}
    \caption{Typical display from the NenuSUN website. The dynamic spectrum of the entire observation for the selected day is displayed on the left. A slider synchronises this spectrum with the 1-minute range images of the four Stokes parameters shown on the right. The ``Context'' tab provides access to observations from other optical or radio observatories, offering complementary information on the solar conditions of the day.}
    \label{fig:NenuSUN-Website}
\end{figure}

 A web interface\footnote{\url{https://nenusun.obspm.fr/catalogue/}} shows the quick-looks of NenuFAR observations. The user visualizes the dates of observation through a calendar, which also provides a quick-look of the entire spectral observation of a selected day. By clicking on the day of interest, the user is directed to a second page where an interactive slider enables selecting the time range for the display of 1-minute windows of the four Stokes parameters (Fig.~\ref{fig:NenuSUN-Website}). Raw data are displayed, i.e., cross-talks for the Stokes parameters are not yet corrected. The ``Context'' tab provides access to a dedicated page that aggregates and displays observations—including images—from various observatories. This feature enhances the interpretability of NenuFAR data by offering a broader, multi-observatory perspective on the observed phenomena.

\subsubsection{NenuSunPy: Dynamic spectra processing tool}
Data access for NenuFAR relies on the NenuPy library, developed by the NenuFAR IT team\footnote{\url{https://nenupy.readthedocs.io}}. This library facilitates the processing of both low-rate and high-rate dynamic spectra (hereinafter TF-data). However, the Python-based methods may present a steep learning curve for non-specialists. Additionally, NenuFAR’s TF-data are exceptionally large, easily exceeding several hundred gigabytes. Consequently, reading an entire observation in one instance is unfeasible. A Jupyter Notebook (named ``NenuSunPy.ipynb'') has been developed to guide the user and generate the commands. The first modules of the tool are designed to streamline information selection for basic processing. Widgets and buttons are proposed to select the time and frequency range (from the display of the low-rate spectrum), to choose the Stokes parameters to process and the cross-talk correction to apply, and to define the time and frequency rebinning to apply. The instructions for NenuPy to read the data are automatically produced and launched from the above information. At this stage, the extracted data can be saved within a `.dill' workspace and that can be downloaded for further analysis on a personal computer. Subsequent modules of the Notebook call Python scripts to perform various tasks, some specifically designed for Type III observations.  \\

\begin{figure}
    \centering
    \includegraphics[width=\linewidth]{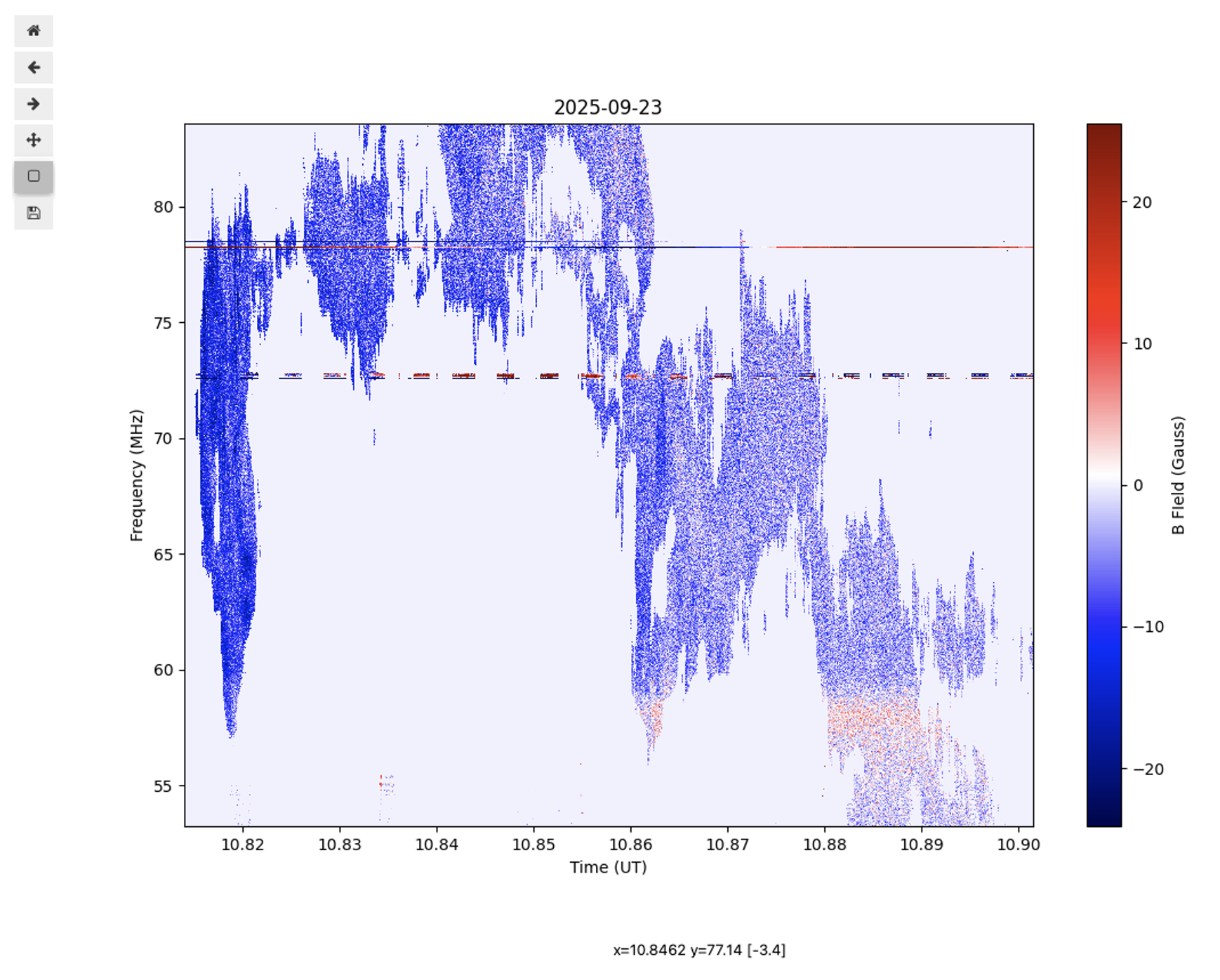}
    \caption{Evaluation of magnetic field during a Type II event observed by NenuFAR: blue and red  area show negative and positive magnetic field, in a time/frequency plane. It also illustrates the widget-orientation of the tool: icons on the top left enable the user to zoom in/out and save the data in png format. }
   \label{fig:Bfield-Radio}
\end{figure}

If a low activity period is available, the user can remove the background of the four Stokes parameters. The method relies on the Lower-Upper technique (LULU) to identify details to remove (or keep) based on their length (in our case, time duration), independently from the signal amplitude \citep{Rohwer2015,Lotz19}. Again, sliders and intermediate displays are provided to guide the choice of parameters, such as period of quiet Sun, window-length for LULU method, and final smoothing. \\
As described in \citet{Cairns09}, any deviation of a electron density model from a $1/r^2$ evolution with distance will produce a spectrum that deviates from a straight line in a (T,1/F) representation. Thus, such a presentation of the dynamic spectrum is proposed.\\
The electron beam velocity responsible for the Type III emission can be inferred from the emission drift rates. When the emission is sufficiently isolated, multiple Type III drift-rates can be extracted using the Hough transform \citep{ZhangP18a}. This method involves three steps, each guided by different sliders with rapid synchronization to the spectrum.  Finally, the user is prompted to choose from several density models to convert the drift rates into electron beam velocities. Our code is an adaptation of the routines provided by \cite{ZhangP18a}\footnote{\url{https://github.com/peijin94/type3detect}}. \\
Time-profile fitting is essential for studying the emission or decay processes and for assessing scattering effects. Second order local maximum techniques detect multiple peaks from the frequency-averaged time profile. Several smoothing methods are proposed to avoid fudge detection. Based on a user-defined threshold, the peaks are detected and the fitting can proceed. Three fitting methods are proposed:  Gaussian, Bi-Gaussian (same peak time, but different standard deviation), and one exponential with two components as proposed by \citet{Chrysaphi24}. An exponential fit is also proposed for the decay phase. Time-profile analysis can be performed on the low- and high-rate data or on the (T,1/F) spectrum.\\
As usual, NenuFAR defines the V/I = (L-R)/(L+R) signal, where L and R refer to Left-handed and Right-Handed polarizations as seen from the observer. When the solar magnetic field is directed towards the observer, V/I is equivalent to the degree of circular polarization $r_c$ = (O-X)/(O+X). So, a positive degree of circular polarization is the signature of a prominent O-mode. \\
The interpretation of Stokes $V$ in terms of magnetic field strength relies on the papers of \citet{Melrose72, Melrose80}. Following assumptions on the alignment of the Langmuir waves with the B-field, the harmonic signal leads to an estimate of the magnetic field strength (Fig. \ref{fig:Bfield-Radio})

Additionally, a dedicated Jupyter Notebook, ``NDASunPy.ipynb'', has been created specifically for NDA data analysis. This Notebook employs the same methodological framework as the NenuFAR tool, utilizes shared analysis utilities, and incorporates data calibration in Jansky based on internal calibration sequences. The most up-to-date versions of the Jupyter Notebooks and related Python routines are available on \url{https://gitlab.obspm.fr/soler/nenusunpy}.

\subsubsection{Radio imaging}
A dedicated Jupyter Notebook, ``Sun\_Imaging.ipynb'', has been developed as part of the NenuFAR SUN imaging workflow \citep{zhangNenuSunIMG}, a modular processing tool developed to produce solar interferometric images from NenuFAR observations in a reproducible way. It is designed to guide users through the main steps of low-frequency solar imaging while keeping the processing chain transparent and repeatable. The workflow is built on standard LOFAR/NenuFAR radio-interferometric software, mainly DP3 for preprocessing and calibration \citep{dp3doc} and WSClean for imaging and CLEAN-based deconvolution \citep{offringaWSClean2014}, together with dedicated modules for solar-specific post-processing and visualisation. In practice, users run the workflow through pre-designed Jupyter Notebook interfaces, supported by accompanying Python modules, where the main parameters and processing steps can be selected interactively. Because the full NenuFAR solar interferometric datasets are very large, the workflow is intended to be run within the Nançay computing environment, including the nancep server, where the raw NenuFAR interferometric data products are available. Access to the data and to the Nançay computing resources needs to be requested and is subject to approval.

The workflow starts from raw NenuFAR Measurement Set for both the solar target and an external calibrator. The main processing chain prepares clean working copies of the calibrator and solar datasets, applies flagging and averaging, selects the solar time range and frequency sub-bands relevant to the event, derives gain solutions from the calibrator, transfers them to the solar data, and finally images the calibrated solar visibilities with WSClean to produce time-resolved FITS image products.

\begin{figure*}
    \centering
    \includegraphics[width=\textwidth]{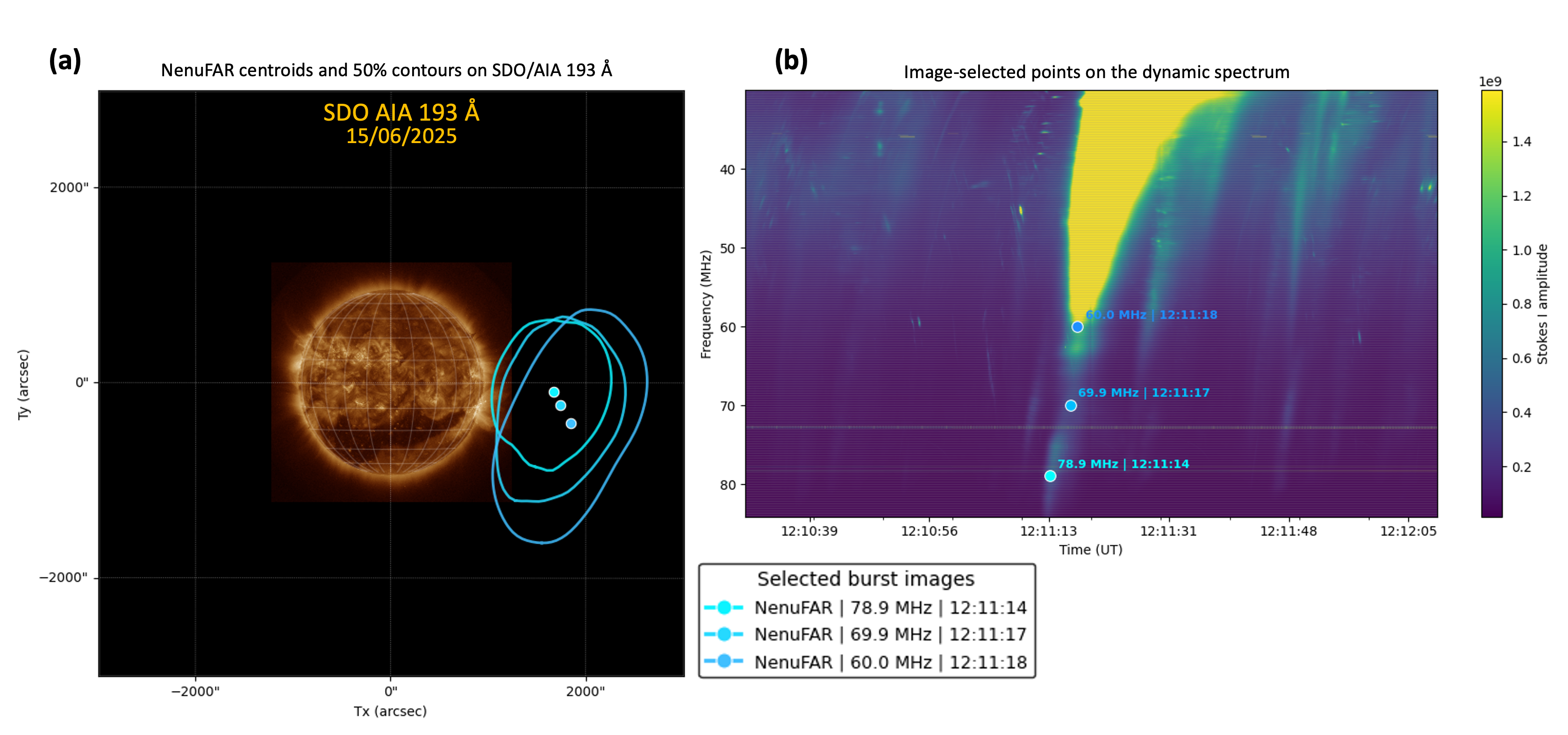}
    \caption{Example post-processing products generated with the NenuFAR SUN imaging workflow. (a) NenuFAR radio-source centroids and 50\% intensity contours measured from selected FITS images and overlaid on an SDO/AIA 193~\AA{} context image. (b) The corresponding image-selected time-frequency points marked on a NenuFAR dynamic spectrum. This output illustrates how the workflow links interferometric source locations with the temporal and spectral evolution of solar radio bursts, supporting source localisation and propagation analyses.}
    \label{fig:NenuFAR_SUN_Imaging_example}
\end{figure*}

Beyond the main calibration and imaging chain, the workflow provides several user-facing tools designed for solar radio-burst analysis. These include quicklook and movie generation from FITS images, instrumental-offset correction using a quiet-Sun reference frame, centroid extraction using two-dimensional Gaussian fitting within a user-defined region of interest, and multi-image source tracking across time and frequency. An example of these post-processing products is shown in Fig.~\ref{fig:NenuFAR_SUN_Imaging_example}: source centroids and 50\% intensity contours measured from selected NenuFAR images are overlaid on an SDO/AIA 193~\AA{} context image, while the same image-selected time-frequency points are marked on the corresponding dynamic spectrum. These modules allow users to connect NenuFAR imaging products directly with dynamic spectra and light curves, making it possible to compare image-selected source locations with the temporal and spectral evolution of radio bursts.

The workflow is modular: each stage can be run independently, so users can repeat only the steps that require refinement without rerunning the full chain. It also supports the processing of multiple sub-bands in a consistent way, which is important for frequency-dependent source localisation and tracking. In this way, the workflow standardises the production of NenuFAR solar FITS images and provides a practical framework for source localisation, quick visual inspection, and burst-propagation studies within the SOLER analysis environment.


\subsection{Coronal modeling}

Modeling methodologies inhabit a key role in supporting the scientific analysis of solar eruptions. Together with accompanying analysis tools, the SOLER modeling tools aim to provide detailed information of the plasma and magnetic field environment at multiple scales from localized regions of interest in the low corona to global coronal scales and further out to establish the connection between the corona and solar wind where in-situ observations are carried out.

\subsubsection{Coronal magnetic field modeling at global scales}

In SOLER, the open-source Python-based coronal magnetic field modeling toolkit CIDER was adopted and further developed for producing routine magnetic field extrapolations at global scales. The toolkit provides two principal modeling methods: the ubiquitous Potential Field Source Surface (PFSS) model as well as a non-potential magnetic field model based on the magnetofrictional (MF) description. In both approaches, the three-dimensional magnetic field  $\mathbf{B}$ is computed in a spherical shell starting from a radius $r=R_0$ up to a given radius $r=R_1$ using
\begin{equation}
    \mathbf{B} = f(r) \nabla \psi
\end{equation}
which results in a Lorentz force
\begin{equation}
    \mathbf{J} \times \mathbf{B} = \frac{1}{\mu_0} \left(\frac{d\ln f(r)}{dr} \right) \left[ -B^2 \mathbf{e}_r + B_r \mathbf{B} \right].
\end{equation}
For the PFSS model, $f(r)=1$ and computing the resulting current-free magnetic field amounts to solving the associated Laplace equation for the scalar field $\psi$ with the boundary condition that the magnetic field becomes purely radial at $r=R_1$, the so-called source surface. On the other hand, the magnetofrictional model seeks to compute a non-potential global magnetic field configuration such that the ideal electric field $\mathbf{E} = - \mathbf{v}_\mathrm{MF} \times \mathbf{B}=0$ where the plasma flow is chosen as $\mathbf{v}_\mathrm{MF} = \frac{1}{\nu}\frac{\mu_0 \mathbf{J} \times \mathbf{B}}{B^2} + v_\mathrm{sw}(r) \mathbf{e}_r$. Given a solar wind outflow velocity, these choices fix the function $f$; $\nu v_\mathrm{sw}(r) = \frac{d\ln f(r)}{dr}$. In contrast to the PFSS model, this model has no set source surface. Instead, the magnetic field becomes increasingly radial with distance due to the prescribed solar wind outflow $v_\mathrm{sw}$.

\begin{figure}
    \centering
    \includegraphics[width=0.75\linewidth]{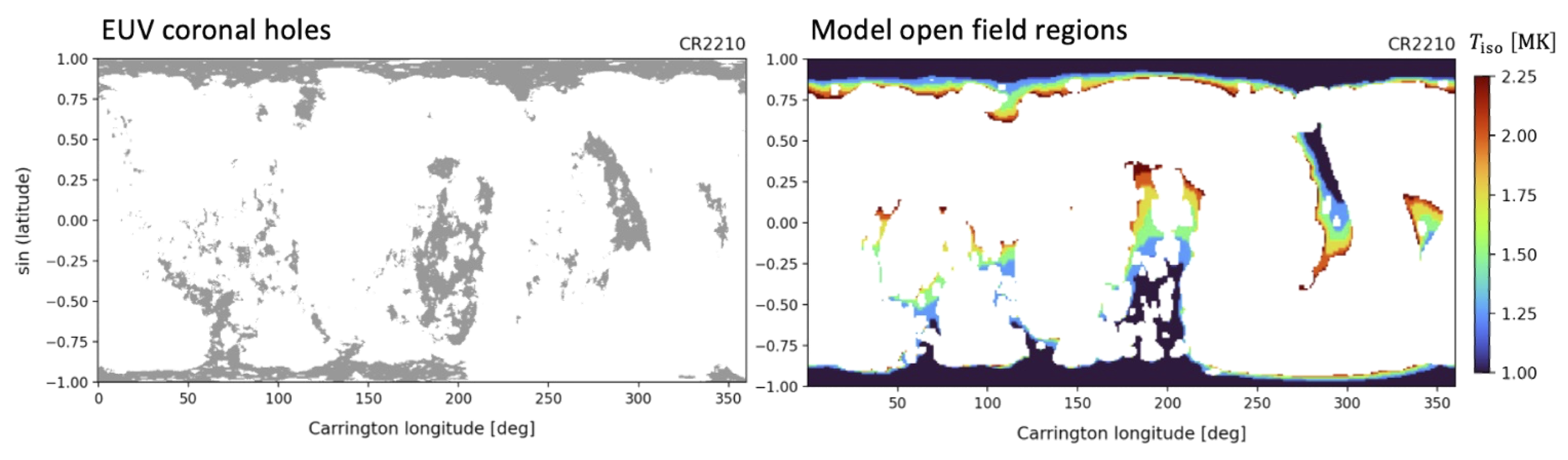}
    \caption{EUV coronal holes and modeled open magnetic field regions. Output generated with CIDER and SunPy.}
    \label{fig:EUV_CH_and_OF}
\end{figure}

The software package CIDER solves these models using an efficient finite difference approach in which the periodic longitude is treated using a spectral decomposition. In addition to the computational routines, the software also includes magnetic map loading and processing utilities as well as functionality related in particular to field line tracing and associated diagnostics. For the magnetofrictional model, various solar wind outflow models are implemented that the user can select and use, including the Parker isothermal solar wind model. 
Due to the use of SunPy, analysis of the magnetic field models can readily be combined with the plotting and visualization capabilities offered by SunPy. An example output of CIDER showing coronal magnetic field maps presented in Fig.~\ref{fig:EUV_CH_and_OF}.

\subsubsection{Physics-informed global nonlinear force-free field extrapolations}

\begin{figure}
    \centering
    \includegraphics[width=\linewidth]{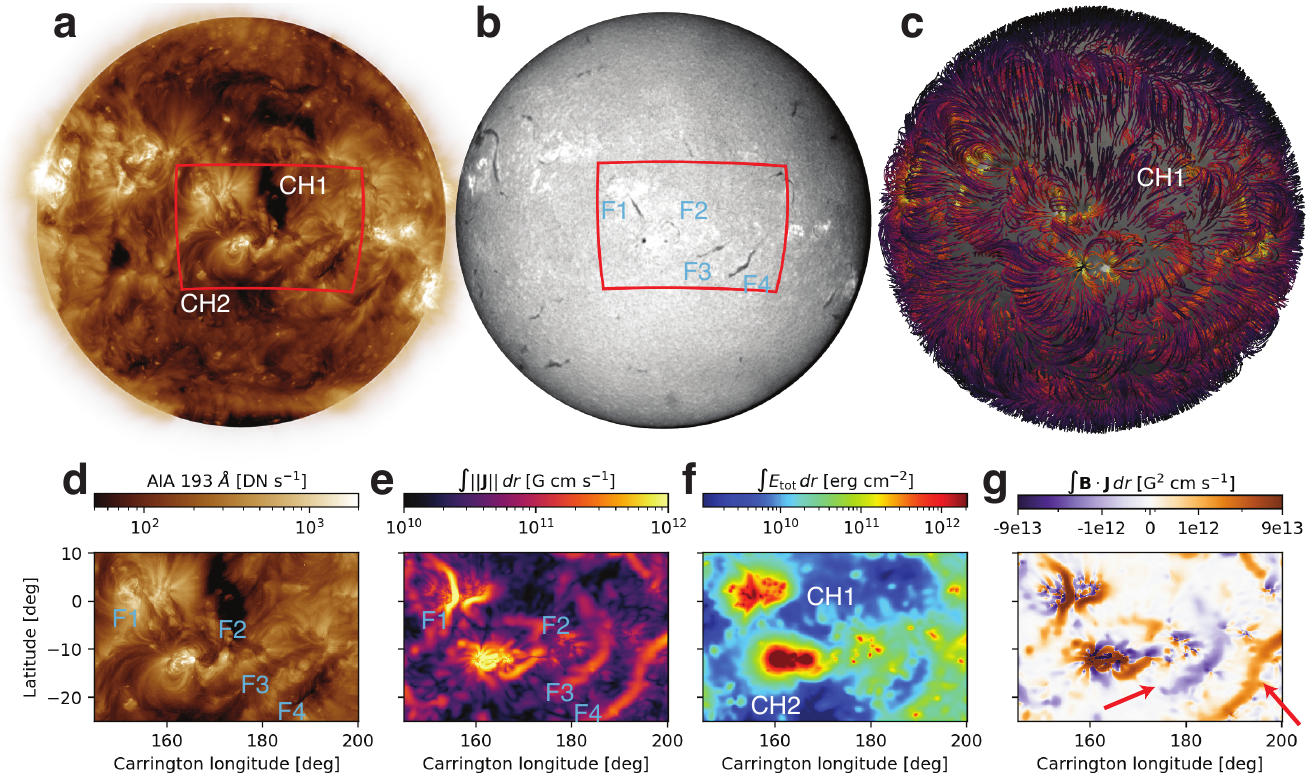}
    \caption{Application of the global PINN-based NLFF extrapolation to the solar magnetic field on 2016 February 5. Panels (a) and (b) show the corresponding full-disk SDO/AIA 193~\AA{} and KSO/H$\alpha$ observations, with filament channels F1--F4 and coronal holes CH1--CH2 indicated. Panel (c) shows magnetic field lines from the extrapolated global NLFF model, color-coded by current density, illustrating active-region connectivity, extended flux ropes corresponding to filament channels, and large-scale open-field structure. Panels (d)--(g) show the subregion marked by the red boxes in panels (a) and (b): AIA 193~\AA{} emission, radially integrated current density, radially integrated magnetic energy density, and radially integrated current helicity, respectively. The extrapolation reproduces filament-channel locations as current-carrying magnetic structures and associates the observed coronal holes with open-field and low-energy regions. The current-helicity map further provides information on the magnetic orientation and handedness of the modeled filament channels (red arrows). This provides a global pre-eruptive magnetic-field context.}
    \label{fig:NF2_global}
\end{figure}

As an additional SOLER activity in global coronal magnetic field modeling, a physics-informed neural network framework was developed for global nonlinear force-free (NLFF) field extrapolations. The tool provides data-driven reconstructions of the pre-eruptive coronal magnetic field under the force-free assumption, using observed vector magnetograms as boundary constraints. Within SOLER, these extrapolations provide a global magnetic context for solar eruptions, linking active regions, filament channels, coronal holes, and large-scale connectivity in a single non-potential magnetic-field model. In particular, the NLFF formulation allows current-carrying magnetic fields, such as those associated with filament channels, to be modeled directly from vector magnetogram boundary constraints, which is not possible in current-free potential-field models such as PFSS. The framework has been released as an open-access software tool \citep{robert_jarolim_2026_20532242}, including Cartesian and spherical/global extrapolation capabilities, with full online documentation and usage examples provided through Read the Docs\footnote{\url{https://nf2.readthedocs.io}}.

Building on the NF2 method introduced by \citet{jarolim2023nf2}, the NLFF extrapolation is formulated as a boundary-value problem in which the magnetic field is constrained by the observed vector magnetic field at the lower boundary and optimized to satisfy the force-free and divergence-free conditions throughout the coronal volume. The PINN acts as a continuous function representation of the magnetic field, mapping spatial coordinates to the three-dimensional magnetic field. By explicitly modeling the vector potential, the method intrinsically enforces solenoidal solutions. This continuous, mesh-free formulation enables NLFF extrapolations from local active-region domains to full-Sun scales, resolving small-scale current-carrying structures while retaining global coronal connectivity. This overcomes key challenges of classical grid-based approaches related to spherical geometry, weak-field regions, and computational cost.

Figure~\ref{fig:NF2_global} shows an application of the method to an SDO/HMI vector magnetogram on 2016 February 5. The extrapolation reconstructs the three-dimensional magnetic field from the low corona to $1.3\,R_\odot$ and captures both local active-region structure and the large-scale organization of the corona. Filament channels visible in EUV and H$\alpha$ observations are reproduced as current-carrying magnetic structures, while coronal holes correspond to open-field regions in the extrapolated field. The derived magnetic diagnostics, including integrated current density, magnetic energy, and current helicity, identify regions of enhanced non-potentiality and provide information on the magnetic orientation of filament channels prior to eruption.

\subsubsection{Probing the coronal Alfv\'en speed and mass density}

The processes responsible for generating energetic particle events as well as emission of electromagnetic radiation are highly dependent on the plasma conditions in the corona. To enable exploration of the prevailing plasma conditions, in SOLER the Solar and Heliospheric Visualization tool (SHELVIS) for analyzing output from state-of-the-art coronal and heliospheric simulations has been developed.
\begin{figure}
    \centering
    \includegraphics[width=0.75\linewidth]{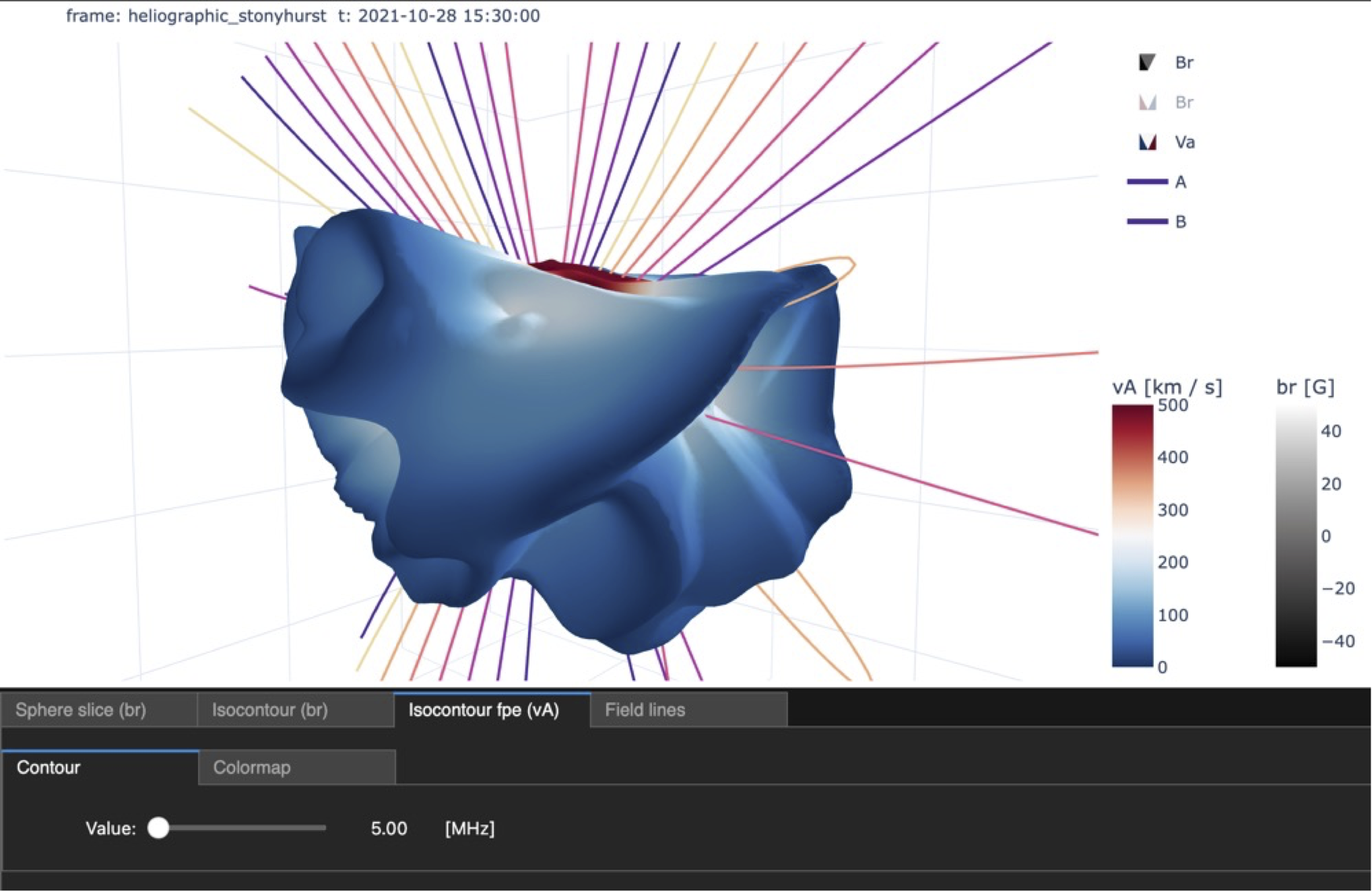}
    \caption{Alfvén speed on a density isosurface from MHD model as determined with SHELVIS.}
    \label{fig:MAST_fpe}
\end{figure}

SHELVIS allows interactive analysis of three-dimensional solar and heliospheric simulation data directly in the Jupyter ecosystem, without requiring familiarity with specialized software traditionally used for such tasks. 
In addition to standard visualization options such as slice and contour plots, SHELVIS also provides easy-to-use complex quantitative analysis capabilities. For instance, the tool allows assessing the spatial distribution of the plasma frequency to aid in estimating possible locations for radio emission.
While a surface of constant mass density can interactively be visualized and explored, the tool moreover allows the surface to be painted by the local value of another quantity such as the Alfvén speed (see Fig.~\ref{fig:MAST_fpe}). Since the plotting is performed in
the notebook, the user has access to all the generated plot data. This not only allows further quantitative analysis, but also opens up the possibility for additional plot capabilities. For instance, tracing magnetic field lines only in regions of low Alfvén speed on
a surface of constant density can be achieved with only a few lines of code.

The primary dataset that is supported and utilized are the openly available simulation outputs from the MAS coronal model provided by Predictive Sciences Inc. Using primarily HMI-based magnetograms as input and relying on a combined physics-based and empirical coronal heating prescription, the model is capable of reproducing the large scale temperature and density structure of the corona \citep[e.g.,][]{Mikic2018}.
It is important to note that SHELVIS is implemented in a model-agnostic way, so that output from other models, e.g. obtained via the VSWMC or CCMC can also be incorporated with minimal effort by the user. In this manner, magnetic field model outputs produced by the CIDER tool can be explored and visualized.

\subsubsection{Geometric reconstruction of CME and shock}

Three-dimensional reconstructions of CMEs and their associated shocks have become essential tools for understanding the physics of solar eruptions. CMEs and shocks are inherently three-dimensional structures whose morphology, kinematics, and interaction with the surrounding corona cannot be fully captured by single-viewpoint or plane-of-sky observations due to the optically thin nature of the plasma. 3D reconstructions allow to more robustly estimate their key physical parameters such as true propagation direction, speed and size, as well as the spatial relationship between the CME flux rope, leading edge, and shock front. This information is critical for constraining models of CME initiation and estimating the initial acceleration and expansion, and for studying the related shock formation and evolution in the low and middle corona. In particular, reconstructing shocks in 3D enables more accurate estimates of shock geometry, Mach number, and obliquity relative to the ambient magnetic field -- quantities that directly control particle acceleration efficiency and the generation of SEP events. 

In SOLER, the open-source Python package CREST was developed to allow geometric reconstructions of CMEs and shocks. 
CREST supports an arbitrary number of remote-sensing imaging data sources to be utilized. The main data type that the user provides as input is a list of SunPy MapSequence objects. MapSequence objects are a sequence of SunPy Map objects, and are designed, for instance, to function as a container for a temporal sequence of remote sensing images. CREST can ingest any MapSequence, agnostic of the observatory or instrument that produced it. The user can input any desired combination of observations via this flexibility. This allows the user, for instance, to use a desired combination of EUV, white-light coronagraph and photospheric magnetic field observations from different observatories. The user interface allows the user to individually select the frame in each of the sequences to view. The images can also be synchronized in time so that the closest image in time of each sequence is selected automatically.

\begin{figure}
    \centering
    \includegraphics[width=0.75\linewidth]{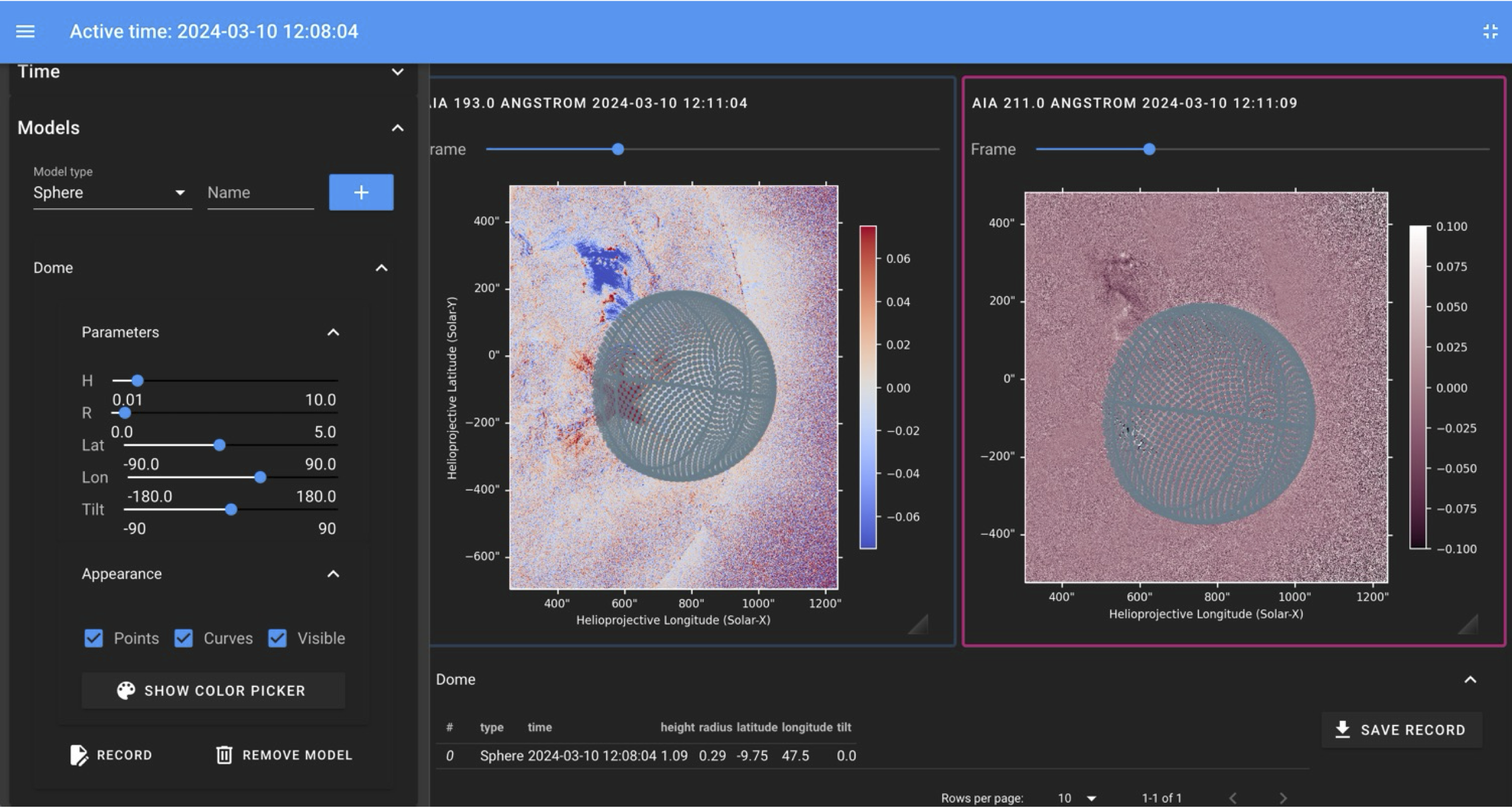}
    \caption{CME and shock geometric reconstruction using CREST.}
    \label{fig:cme_reconstruction}
\end{figure}

As CREST take any MapSequence as input, the user may create the input data in their preferred manner. The typical use-case is to employ the existing rich methods available in SunPy for downloading and processing imaging observations. The Jupyter ecosystem provides a powerful environment for managing this step. However, as an alternative, a GUI that allows to easily browse and download available imaging data using the Helioviewer API has been implemented as part of CREST. This on-line data source provides near real-time remote sensing imagery that have undergone several standardized processing steps before being stored in the publicly accessible server. Thus, the Helioviewer service provides ready-to-use efficient datasets for many use-cases utilizing imaging observations. The implemented GUI follows the intuitive interface familiar to users of the Helioviewer website \url{https://helioviewer.org} or stand-alone application. By using the built-in GUI, a large selection of imagery can be obtained without any detailed knowledge of data acquisition with SunPy. Figure~\ref{fig:cme_reconstruction} shows the CREST user interface. 

CREST currently implements three basic geometric objects: the sphere, the ellipsoid and the GCS model \citep{Thernisien2006, Thernisien2011}. The user can include an arbitrary number of any of these models when carrying out their analysis. The user interface allows to flexibly add, modify and remove models in addition to interactively changing the parameters of the models. The parameters of any model can be recorded at any time and saved for further analysis.

In addition to the aforementioned symmetric geometric models, CREST implements options for deforming the models. The implemented approach uses the free-form deformation technique used in computer graphics applications that efficiently and flexibly allows to deform any object. Although the method is highly flexible, a restricted approach has been implemented in order to manage the complexity of the user interface required to use the method. Thus, currently only deformations of a single plane of the rectangular cuboid that forms the deformation hull of the model at a time is supported. 

A key advantage of the Jupyter Notebook/Lab environment used together with SunPy is that the user can perform any desired preprocessing steps to the imaging data before use in the extended analysis. For example, the user can apply image processing to enhance certain aspects of the images using SunPy (together with other packages, such as the affiliated package sunkit-image) and then provide the result to CREST. However, the GUI also includes interactive image processing options for common operations. In particular, running-difference as well as base-difference images can be created using this option. The interface allows the user to select the reference image to use, as well as the offset between frames to use for the case of running differences. In addition, interactive control of the colour map is implemented.

\subsubsection{Tools for constructing solar photospheric magnetic and electric field maps}\label{sec:electricit}

A key aspect in modeling the corona is the requirement of a map of the magnetic field in the low corona or photosphere to be provided as input. While ready-to-use datasets produced by various observatories are frequently employed by the community, the available data products do not cover all uses cases and requirements of contemporary models. For instance, in cases where several active regions are located nearby, the Spaceweather HMI Active Region Patch (SHARP) cutouts are often split into distinct partially overlapping patches, making their use in a larger domain cumbersome. On global scales, the low cadence of the standard global maps exclude their use in time-accurate modeling. Furthermore, many data products are built from the LOS component, with the full vector field information provided, e.g., by HMI not taken advantage of.  

To address these needs, a Python-based software package, the Electric Field Inversion Toolkit (ELECTRICIT) was developed and in SOLER modernised and rewritten to extensively utilize SunPy functionality, in particular as pertains to the use of the World Coordinate System (WCS). The software allows the user to flexibly create custom magnetic map datasets based on exploiting HMI full disk vector magnetograms. In addition to isolated cutout patches both at global and local scales, methods for combining custom
cutouts with synoptic or synchronic maps have been developed. This allows for creating custom maps that combine high-resolution local areas with global maps, providing new capabilities especially for modeling schemes that support refined local meshes. ELECTRICIT also features functionality for constructing maps of the photospheric electric field, which plays a crucial role as the driver for the transport of magnetic energy and helicity into the corona. Constructing such maps requires
an inversion of Faraday’s law based in particular on the observed time-series of vector magnetograms. The implemented methods are detailed in \citet{Lumme2017}.

\subsubsection{Electron shock acceleration modeling}\label{sec:easi}

SOLER provides a Python-based simulation code to model the acceleration of electrons at coronal shocks. The code, called Electron Acceleration SImulations (EASI), has been described in a recently published paper \citep{nyberg2026} and has been released as open-source software \citep{easi_2026}. The code computes the steady-state distribution function of electrons in the vicinity of shocks, modelled as a transition of finite thickness (instead of the typically assumed step-like shock). The code traces the electrons in a large-scale magnetic field that has a hyperbolic tangent profile across the shock with a flow profile that obeys the conservation laws (mass, momentum and energy) across the shock. The effect of turbulent fluctuations is modelled using a stochastic approach, i.e., performing random scatterings with a mean free path proportional to the intensity of turbulence. We use a model of mean free path that consists of an ambient component present over the whole simulation domain plus a separate strong turbulence component that is present only in the shock transition, mimicking instabilities at the shock. The model thus encompasses three acceleration mechanisms: diffusive shock acceleration, shock drift acceleration, and stochastic shock drift acceleration, and their relative roles can be controlled by the user changing the mean free path parameters. The model is able to provide a realistic description of electron acceleration in coronal shocks and also assess the parameter range that could produce electron beams able to produce type II bursts.

\subsection{Magnetic connectivity analysis}

\subsubsection{Coronal magnetic field analysis with CIDER}

\begin{figure}
    \centering
    \includegraphics[width=0.75\linewidth]{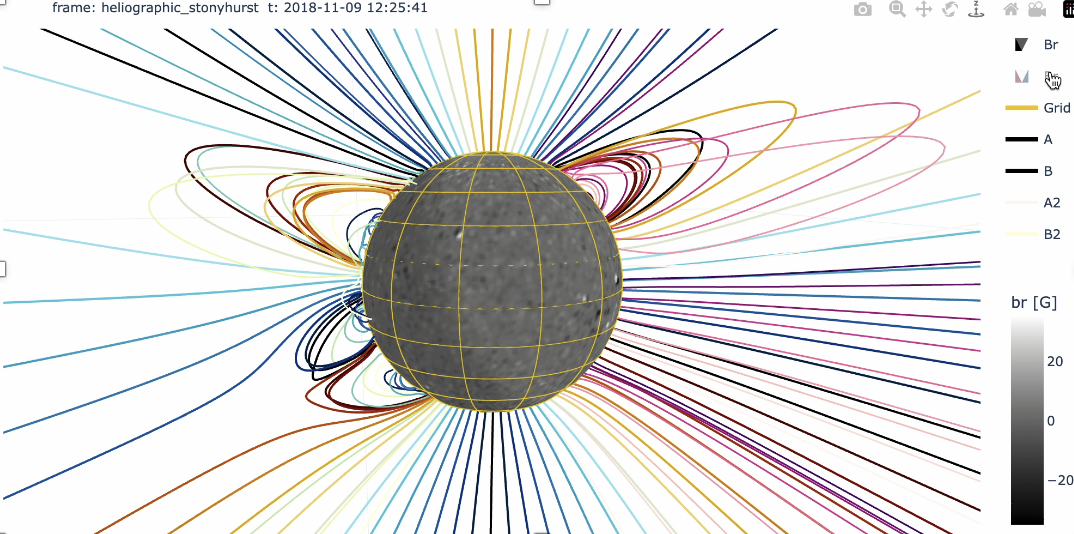}
    \caption{Global magnetic field topology provided by CH-optimized modeling using 
    (CIDER and SHELVIS.}
    \label{fig:MF}
\end{figure}

Estimating the magnetic connectivity between structures and regions in the solar corona as well as the connection from the corona to the measuring spacecraft is crucial for assessing the source regions responsible for the measured energetic particles. The complex connectivity in the solar corona can be estimated using the models and methods implemented in CIDER and SHELVIS, which is illustrated in Fig.~\ref{fig:MF}. Using CIDER, the magnetic field configuration provided by the magnetofrictional outflow model can be employed, in addition to the PFSS model, to establish connectivity between the lower and upper corona. In addition, using the analysis tools provided by SHELVIS, the connectivity as estimated using MAS model outputs can be computed. 

\subsubsection{PFSS extension of Solar-MACH}
The Solar MAgnetic Connection Haus (Solar-MACH) tool \citep{Gieseler2023} provides ephemeris and magnetic connection information as well as visualization for the heliospherc spacecraft fleet through the Python package solarmach \citep{solarmach_2026} and a Streamlit\footnote{\url{https://pypi.org/project/streamlit/}} interface \citep{solarmach_streamlit_2023}.
\begin{figure}
    \centering
    \includegraphics[width=0.75\linewidth]{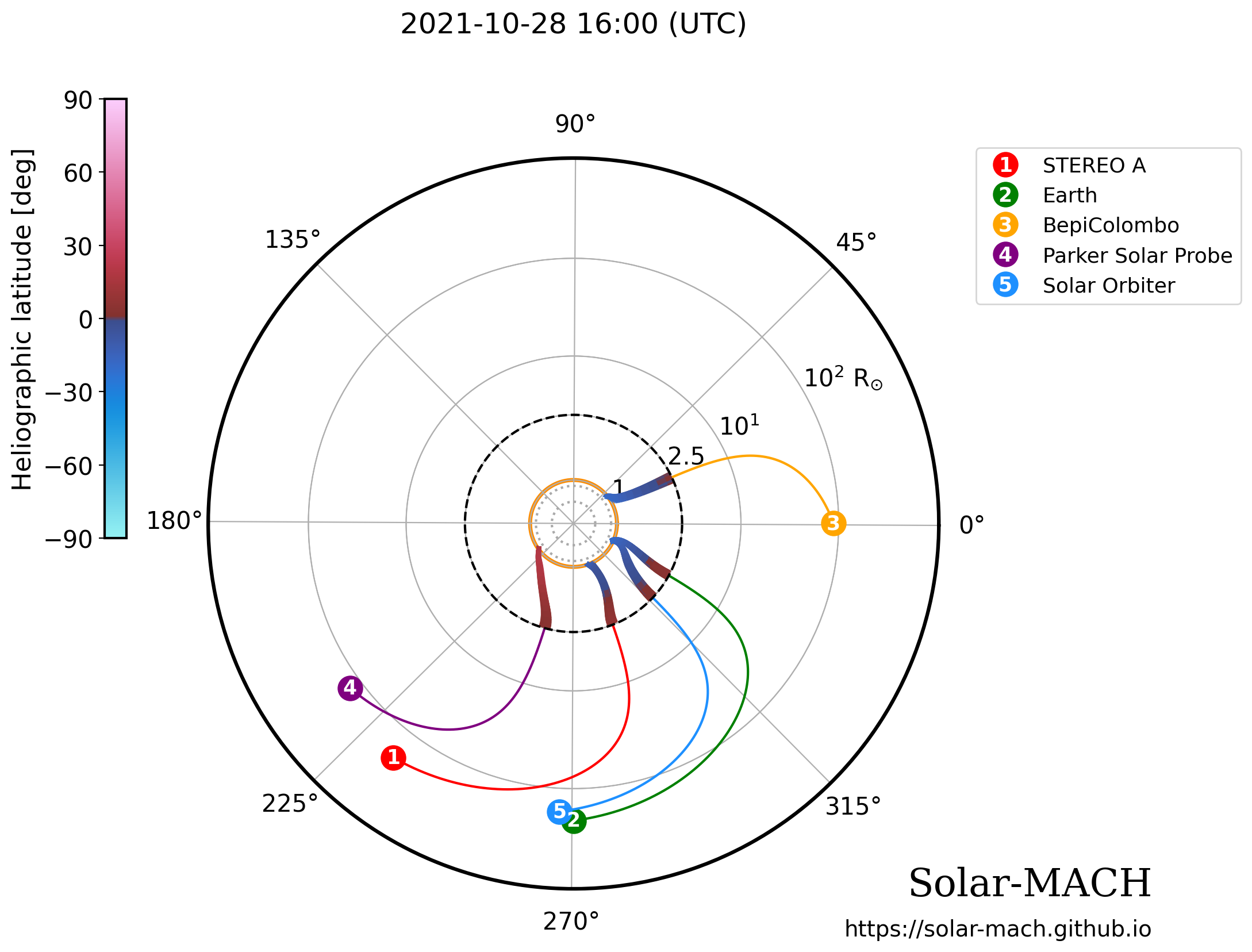}
    \caption{Solar-MACH's PFSS extension that extrapolates the magnetic connectivity between the Sun and an observer in two distinct segments. From 1 to 2.5 solar radii (chosen source surface height), a PFSS model is used and the figure is drawn on a linear scale. Beyond the source surface, ballistic backmapping is applied and the figure uses a logarithmic scale.}
    \label{fig:solarmach_pfss}
\end{figure}
Solar-MACH's PFSS extension extrapolates the magnetic connectivity between the Sun and an observer in two distinct segments, as shown in Fig.~\ref{fig:solarmach_pfss}. Within the interplanetary space, the magnetic connection is derived using a classic Parker spiral approximation. Close to the Sun -- inside a sphere that's defined by the height of the source surface -- a Potential Field Source Surface \citep[PFSS; see][and references therein]{Mackay2012} model of the magnetic field is employed instead. By default, the source surface is set to 2.5 solar radii, but this is an adjustable parameter. 
The PFSS solution is calculated using a magnetogram \citep[in form of an automatically obtained GONG map;][]{Harvey1996}, which defines the lower boundary condition for the unique magnetic field solution. It is then used to find the magnetic connection between the point at which the Parker spiral approximation ends on the source surface and the photosphere. Optionally, instead of a single point on the source surface, a bundle of field lines may be traced starting from a ring of equidistant points at the source surface centered around the point at which the observer's Parker spiral intersects with it.

\subsection{SEP analysis}
The SEP analysis tools of the SOLER project are largely built upon the data loading infrastructure of the Python package SEPpy \citep{seppy_2026} and further methodology developed within the SERPENTINE project\footref{SERPENTINE} \citep{Palmroos2022} that has also been integrated into SEPpy. If possible, existing Python packages such as solo-epd-loader \citep{solo-epd-loader_2025}, speasy \citep{speasy_2025}, STIXpy \citep{stixpy_2026}, or sunpy \citep{sunpy_community2020, sunpy_2026} are used. SEPpy supports energetic particle measurements (focusing on electron and proton observations) by BepiColombo, JUICE, Parker Solar Probe, SOHO, Solar Orbiter,  STEREO~A, and Wind. Table~\ref{tab:spacecraft_instruments} gives an overview of the instruments currently supported in the different tools. All tools are made available as Jupyter Notebooks through a single repository \citep{sep_tools_2026} and are effortlessly available through SOLER's JupyterHub server ({see Sect.~\ref{sec:hub}}).

\begin{table}[t]
\centering
\newcommand{\y}{\makebox[1em]{\checkmark}}
\newcommand{\n}{\makebox[1em]{--}}
\newcommand{\x}{\makebox[1em]{$\times$}}
\caption{Energetic particle instruments supported in the different SEP analysis tools. 
\y~indicates supported, \n~not supported (yet), and \x~not possible/planned.
}
\label{tab:spacecraft_instruments}
\begin{tabular}{rl *{7}{c}}
\toprule
\textbf{Spacecraft}\rule{0pt}{3.5cm} & \textbf{Instrument} &
\rotatebox[origin=lb]{90}{\textbf{Multi-Spacecraft-Plot}} &
\rotatebox[origin=lb]{90}{\textbf{Multi-Instrument-Plot\vphantom{y}}} &  
\rotatebox[origin=lb]{90}{\textbf{PyOnset}} &
\rotatebox[origin=lb]{90}{\textbf{Regression-Onset}} &
\rotatebox[origin=lb]{90}{\textbf{Spatial-Distribution}} &
\rotatebox[origin=lb]{90}{\textbf{PAD-and-Anisotropy}} &
\rotatebox[origin=lb]{90}{\textbf{Spectra}} \\
\toprule
BepiColombo                        & SIXS-P                            & \y                  & \n & \y & \y & \n & \n & \n \\ \midrule
JUICE                              & RADEM\tablefootmark{a}            & \y & \n & \n & \n & \n & \x & \n \\ \midrule
\multirow[t]{2}{*}{PSP}            & ISOIS/EPI-Lo\makebox[0pt][l]{\tablefootmark{b}}     & \y                  & \y & \y & \y & \n & \x & \n \\
                                   & ISOIS/EPI-Hi/HET\makebox[0pt][l]{\tablefootmark{b}} & \y                  & \y & \y & \y & \y\makebox[0pt][l]{\tablefootmark{c}} & \x & \y\makebox[0pt][l]{\tablefootmark{c}} \\ \midrule
\multirow[t]{3}{*}{Solar Orbiter}  & EPD/STEP                          & \n                  & \n & \y & \y & \x & \x & \n \\
                                   & EPD/EPT                           & \y                  & \y & \y & \y & \n & \y & \y \\
                                   & EPD/HET                           & \y                  & \y & \y & \y & \y\makebox[0pt][l]{\tablefootmark{c}} & \y & \y \\ \midrule
\multirow[t]{2}{*}{STEREO A \& B}  & SEPT                              & \y                  & \y & \y & \y & \n & \y & \y \\
                                   & HET                               & \y                  & \y & \y & \y & \y\makebox[0pt][l]{\tablefootmark{cd}} & \x & \y \\ \midrule
\multirow[t]{2}{*}{SOHO}           & EPHIN                             & \y                  & \y & \y & \y & \x & \x & \n \\
                                   & ERNE                              & \y                  & \y & \y & \y & \y\makebox[0pt][l]{\tablefootmark{c}} & \x & \y \\ \midrule
Wind                               & 3DP                               & \y                  & \y & \y & \y & \n & \y & \y \\
\bottomrule
\end{tabular}
\tablefoot{
    \tablefoottext{a}{Only available as count rates (so far).}
    \tablefoottext{b}{Electrons only available as count rates (so far).}
    \tablefoottext{c}{Only protons (so far).}
    \tablefoottext{d}{STEREO B not supported.}
}
\end{table}

\subsubsection{SEP intensity-time profiles as observed by multiple spacecraft}
\label{sect_Multi-Spacecraft-Plot}
\begin{figure}
    \centering
    \includegraphics[width=1.0\linewidth]{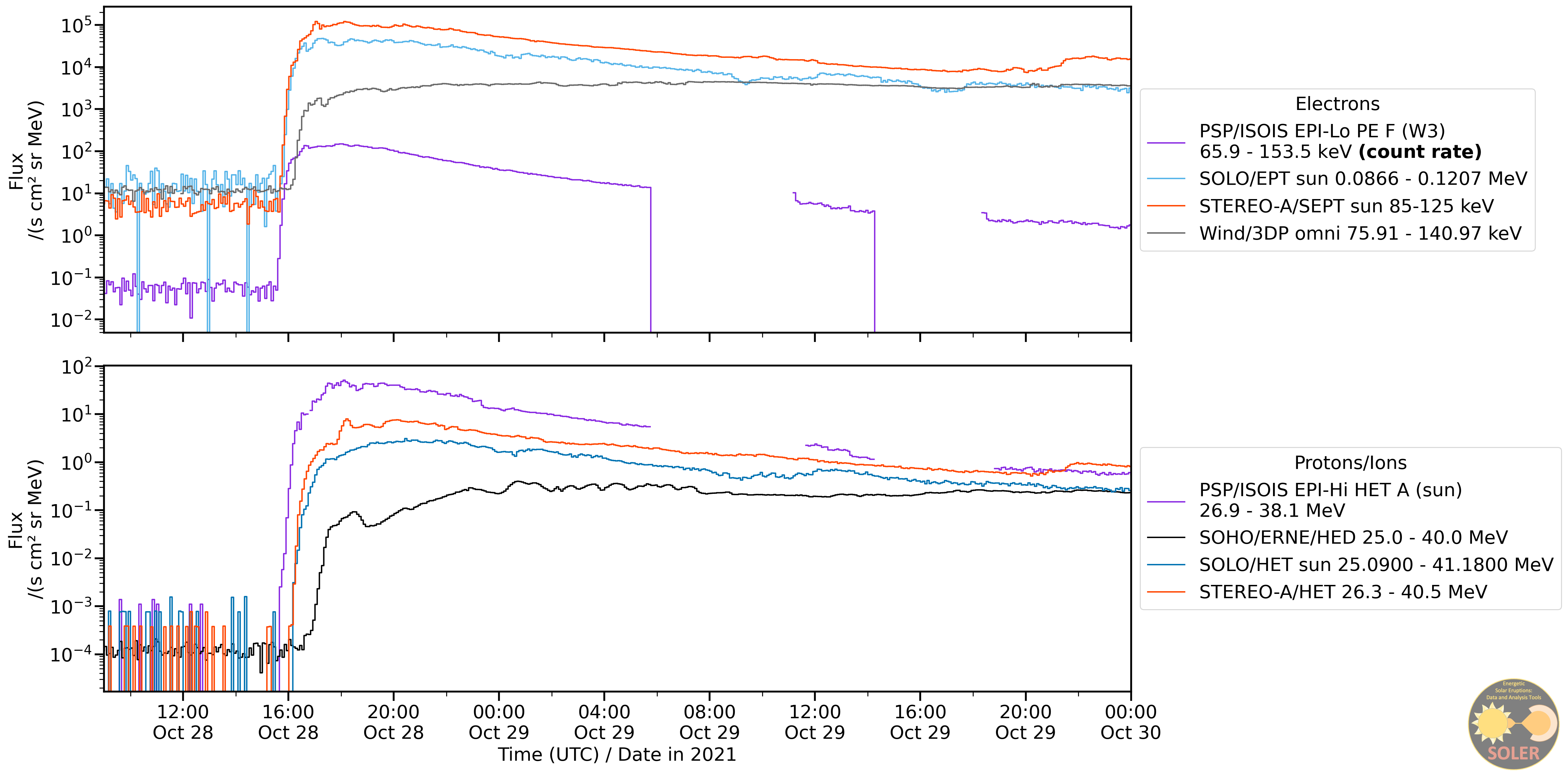}
    \caption{Output of the Multi-Spacecraft-Plot tool for the 28th October 2021 event showing fluxes of different spacecraft-instrument combinations for $\sim$100~keV electrons (top) and 25-40~MeV protons (bottom).}
    \label{fig:sep_multi-sc-plot}
\end{figure}

The \textbf{Multi-Spacecraft-Plot} tool allows visualizing energetic particle data of various instruments onboard different spacecraft together in a single plot per species, enabling direct comparison of the event across the heliospheric spacecraft fleet. Figure~\ref{fig:sep_multi-sc-plot} shows such a resulting plot, which allows for example to compare intensity levels and timing of the event as detected at different locations.

The tool provides a GUI to select electron and proton (ion) measurements by different spacecraft and instruments. After defining the time range, the user can select the viewing directions of applicable instruments and is provided with tables defining the energy channels of the different instrument.
This consequently allows
to select comparable energies across multiple spacecraft-instrument combinations, with the option to combine adjunct energy channels for some instruments. Further options include changing the line colors, resampling of the data, or adjusting the plotting range in time. Furthermore, the Notebook offers example code to manually edit the figures after creation, which can be useful to add features such as vertical lines or shadings.
To further support the comparison between differently located observers, a corresponding Solar-MACH plot and table can be easily created at the end of the Notebook.

\subsubsection{SEP time profiles and their relation to interplanetary context}
\begin{figure}
    \centering
    \includegraphics[height=0.7\textheight]{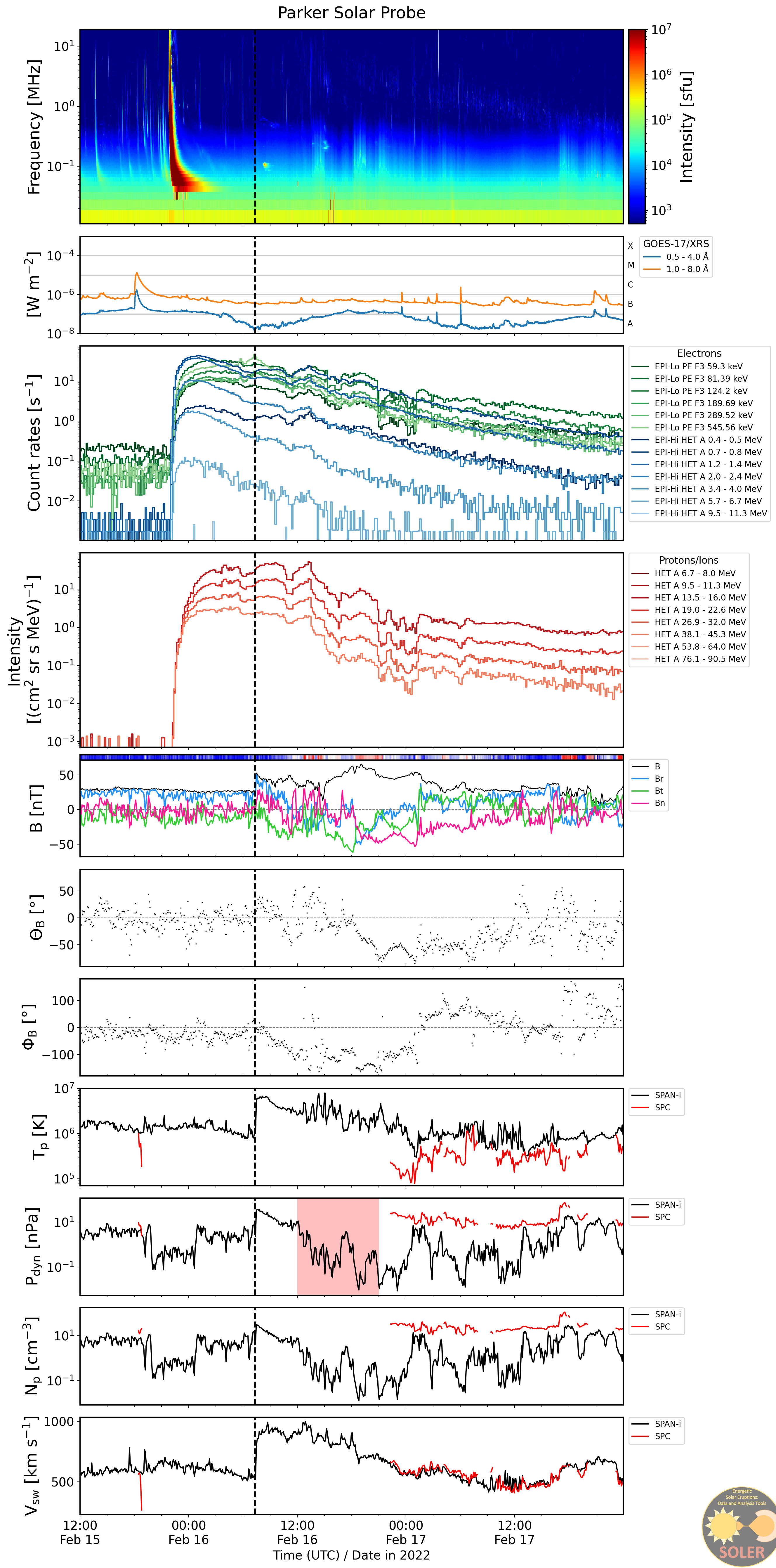}
    \includegraphics[height=0.7\textheight]{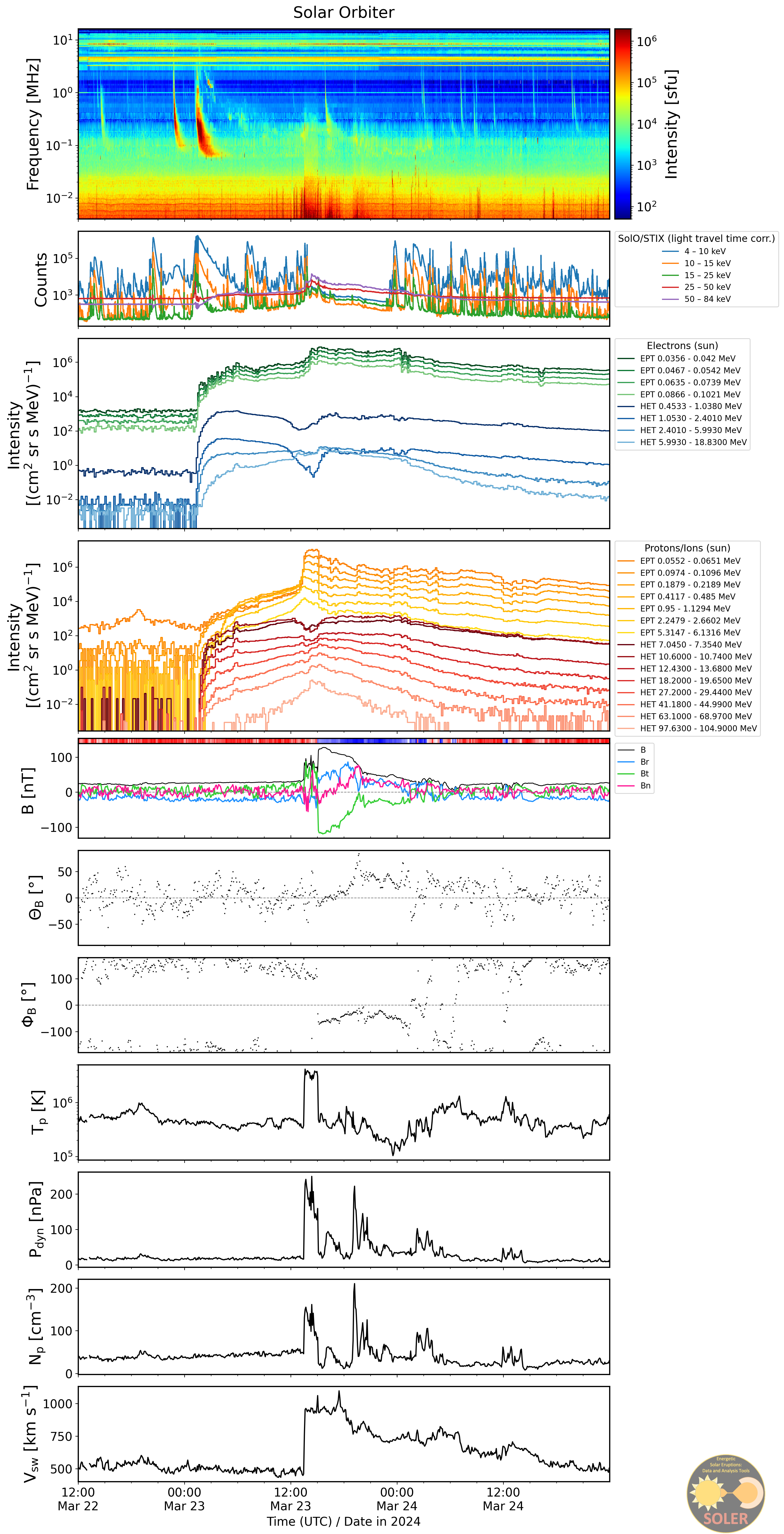}
    \caption{Output of the SEP-Multi-Instrument tool for two different events showing radio, X-ray, charged particles, magnetic field, and plasma observations by PSP (left; with GOES for X-Ray) and Solar Orbiter (right). The left figure has been edited after initial plotting via the tool to add a vertical line and a red-shaded region.}
    \label{fig:multi-instrument}
\end{figure}

Similar to the Multi-Spacecraft-Plot tool (\ref{sect_Multi-Spacecraft-Plot}), the \textbf{Multi-Instrument-Plot} tool provides a multi-panel figure, but this time not for different spacecraft, but different instruments onboard a single spacecraft. Figure~\ref{fig:multi-instrument} gives examples for PSP (left) and Solar Orbiter (right). Multiple energy channels of charged particle observations can be put into their interplanetary context, including radio observations, X-ray, magnetic field, and plasma measurements. For comparison, the X-ray observations of GOES can always be added. Similarly to the previous tool, a sophisticated GUI allows for most of the necessary options to be set, including defining viewing direction and energy channel selection for charged particle detectors.
Finally, the notebook also offers examples on how to edit the figure after creation, such as changing the order of panels or adding vertical lines and shadings (cf. Fig.~\ref{fig:multi-instrument}, left).

\subsubsection{SEP onset determination}

The \textbf{PyOnset} \citep[][]{Palmroos2025, Palmroos_Gieseler2025} and \textbf{Regression-Onset} tools are two SOLER onset determination tools that work on different operational principles.

PyOnset employs a novel combination of a modified Poisson-CUSUM scheme, which is coupled with statistical bootstrapping and methodological time-averaging. It is designed to find the most probable onset time regardless of the time resolution, and provide an uncertainty related to the onset time. The software makes it easy to determine onset times in a selection of energy channels of a particular instrument, and conveniently apply either time-shift analysis (TSA) or velocity dispersion analysis (VDA) on the onset times to infer the particle injection times at the Sun \citep[e.g.,][]{Vainio2013}. The accompanied publication \citep{Palmroos2025} gives more details.

\begin{figure}
    \centering
    \includegraphics[width=0.8\linewidth]{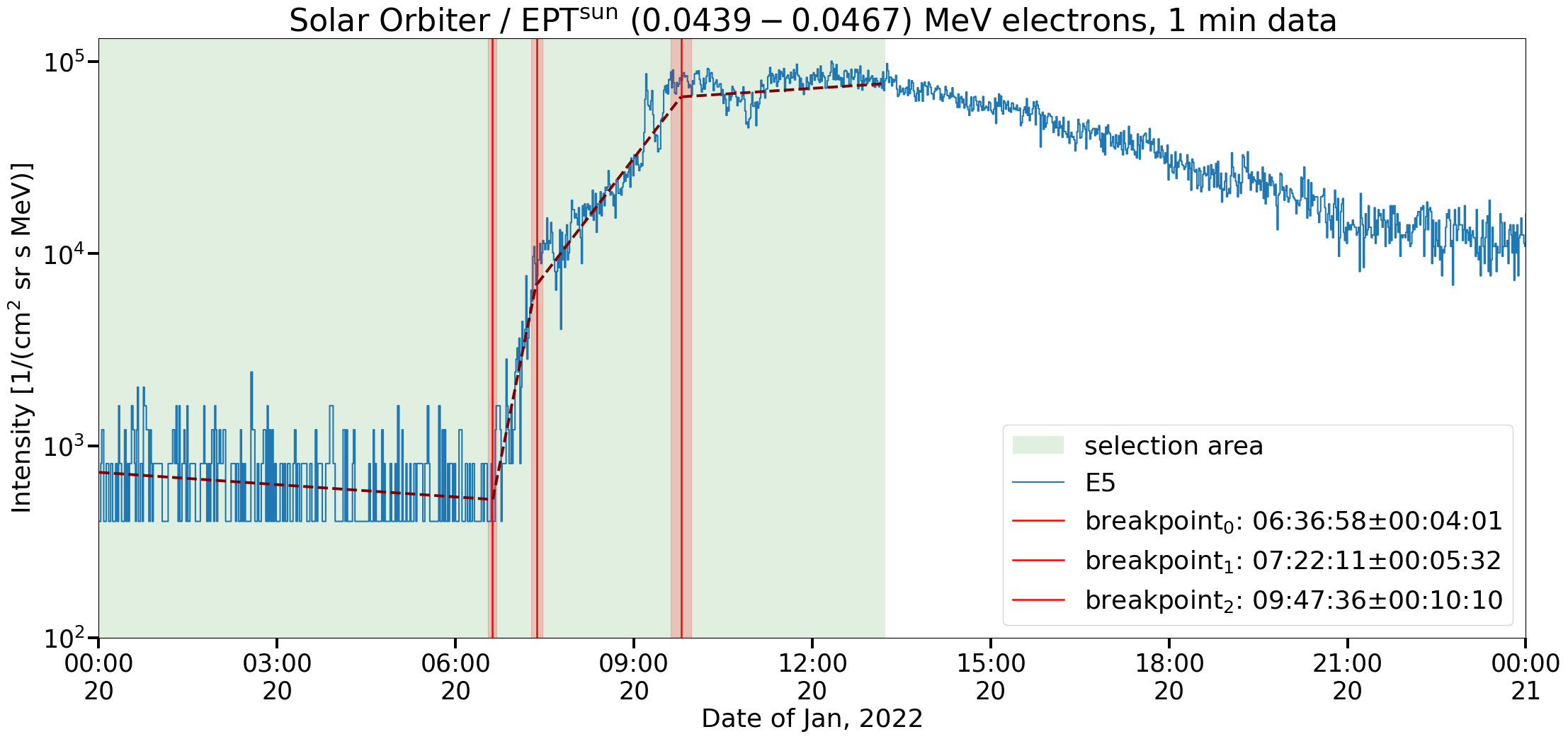}
    \caption{The SEP-Regression Onset Tool finds breakpoints in intensity time series data, which are marked with vertical red lines and listed in the legend. The uncertainty related to the position of the breakpoint is indicated with partly red shading, and also listed in the legend. The green shading on top of the time series indicates the selection of the data that is inspected by the tool.}
    \label{fig:linreg}
\end{figure}

The Regression-Onset tool applies segmented linear regression to the logarithm of intensity to find breakpoints where a linear trend changes from one to another. One such breakpoint can be taken to be the onset of an SEP event. Figure \ref{fig:linreg} illustrates this with Solar Orbiter's EPT 439--467 keV electron intensity time series, where three distinct breakpoints have been found in the data. The first breakpoint corresponds to the onset of the event, while the latter two breakpoints coincide with changes in the trend of the intensity time series.

\subsubsection{Longitudinal SEP distributions and their temporal evolution}

The \textbf{SEP-Spatial-Distribution} tool makes use of the observer-dependent intensity-time series observations to characterize the overall particle distribution in space in terms of its time-dependent width and center using Gaussian fits. 
\begin{figure}
    \centering
    \includegraphics[width=0.85\linewidth]{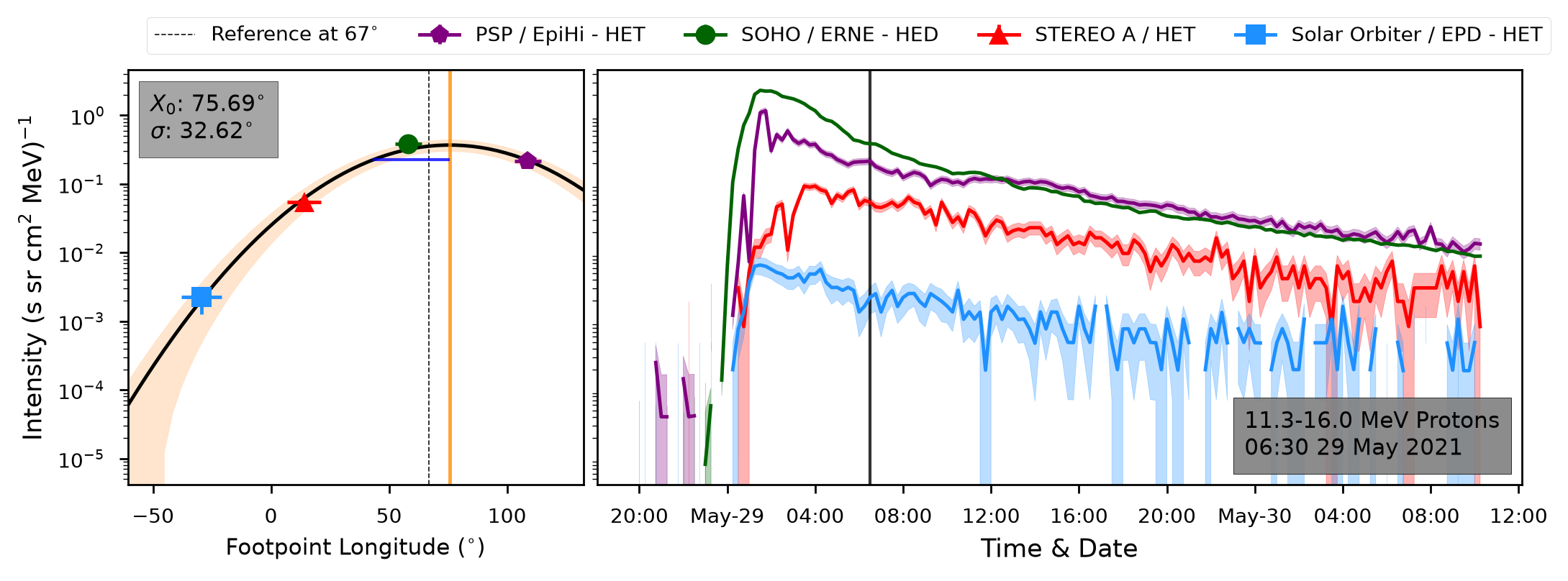}
    \caption{Gaussian fit (left) applied to the intensity measurements at the time marked by the vertical line in the right-hand panel, which shows the time series of the multi-spacecraft observations of $\sim14$~MeV protons during the 28 May 2021 event. The Gaussian parameters are distinctly presented with an orange vertical line for $X_0$ and a horizontal blue line for $\sigma$.}
    \label{fig:sep_distribution_gauss}
\end{figure}
\begin{figure}
    \centering
    \includegraphics[width=0.75\linewidth]{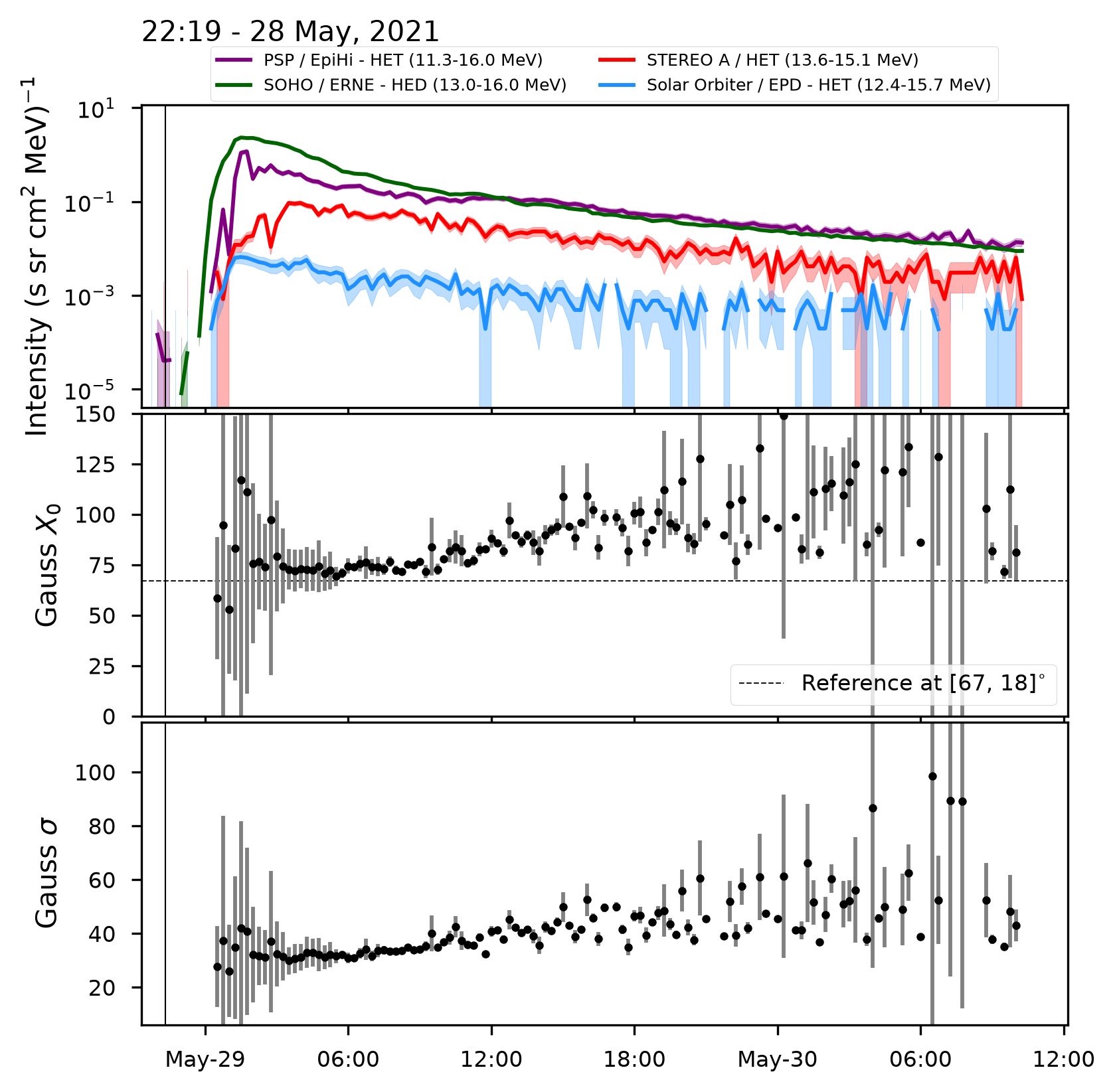}
    \caption{Temporal evolution of the longitudinal distribution of $\sim$14~MeV protons during the 28 May 2021 {SEP} event, which is associated with a solar flare at 22:19 UT (time marked by the black vertical line and flare longitude marked in the center panel with a dashed horizontal line). Top: Intensity-time profiles measured by different spacecraft; middle and bottom: Fit results of the longitudinal intensity distributions, Gaussian $X_0$ and $\sigma$.}
    \label{fig:sep_distribution_final}
\end{figure}
It is extending the work by \citet{Dresing2018} and \citet{Kahler2023} by employing, for the first time, a fleet of spacecraft consisting of more than three observers, which allows us to constrain a Gaussian function.

After choosing a date, the tool loads energetic particle data measured by multiple observers as well as their positional data. 
Then a Solar-MACH plot of the spacecraft constellation is shown, followed by an SEP time-series plot. 
The user chooses an interval for background subtraction, then radial scaling, as well as intercalibration factors, can be applied.
Gaussian curves are fitted to the longitudinal intensity distributions, as shown in Fig.~\ref{fig:sep_distribution_gauss}. 
The curves are defined using $y=A\cdot\exp\left[-\frac{(x-X_0)^2}{2 \sigma^2}\right]$, where $A$ is the maximum intensity, $X_0$ is the center of the distribution, and $\sigma$ is the standard deviation (used as a measure of the width) of the distribution.
These fits are applied to time slices throughout the event so that the Gaussian parameters can be determined as a function of time. 
The final output of the tool is a time-series plot of the time-dependent Gaussian parameters ($X_0$ and $\sigma$) together with the SEP intensity-time series (Fig.~\ref{fig:sep_distribution_final}).

As can be seen in the figure, the center of the Gaussian distribution begins to deviate from the recorded flare longitude (dashed line in the center panel) after around 06:00 UT on 29 May 2021. 
This deviation to western longitudes is in agreement with the direction one would expect from a {CME}-driven shock source, which moves away from the Sun and therefore consecutively connects to more western longitudes. 
The SEP-Spatial-Distribution tool, therefore, provides a powerful new method to study the temporal evolution of SEPs in the inner heliosphere and to disentangle different source processes.

The tool is designed to study 14~MeV ($11-16$~MeV) proton datasets, with corresponding radial scaling and intercalibration factors. 
The user is able to change the proton energy channels loaded from each instrument, but the radial scaling and intercalibration factors must also be changed to match the new ranges.
The incorporation of electron datasets has not yet been introduced.

\subsubsection{SEP pitch-angle distributions and first-order anisotropies}
\begin{figure}
    \centering
    \includegraphics[width=0.75\linewidth]{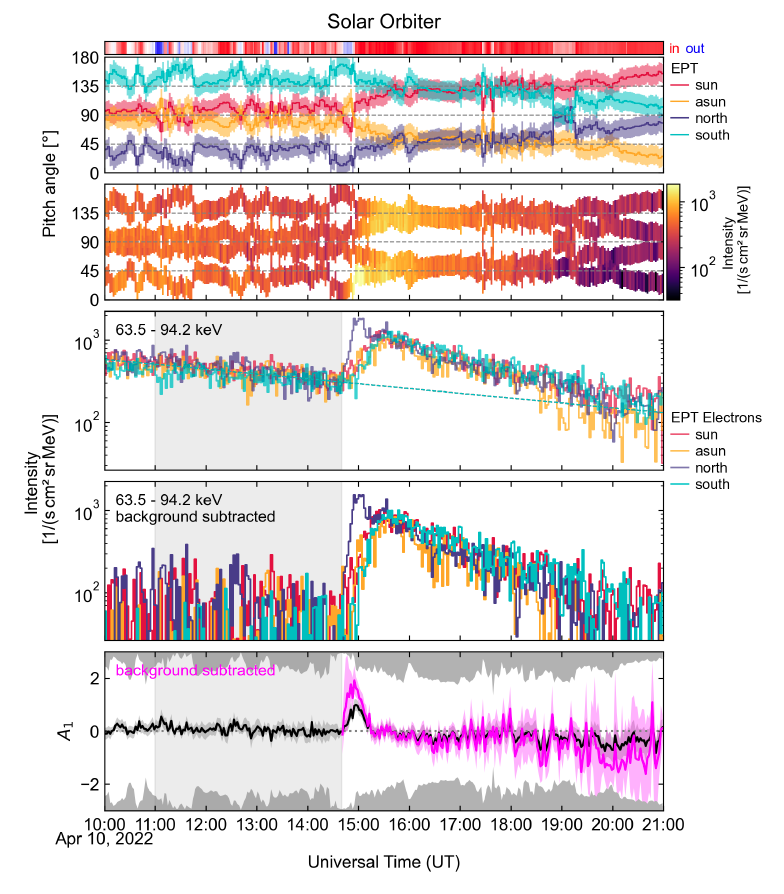}
    \caption{Output plot of the {SEP}-{PAD}-and-Anisotropy tool for the solar energetic electron event of 10 April 2022. From top to bottom: Pitch-angle ranges covered by the four viewing directions of  {SolO} /  {EPT}, pitch-angle distribution with electron intensities marked by color coding, intensity-time series of 64--94 keV electrons observed in the four viewing directions, the same intensity-time series but with background subtraction applied (gray shade marks the background interval), and first-order anisotropies both with (magenta) and without (black) background subtraction.}
    \label{fig:sep_pad}
\end{figure}
The \textbf{{SEP}-{PAD}-and-Anisotropy} tool determines and visualizes energetic particle pitch-angle distributions (PADs) and allows the user to calculate the first-order anisotropies from these PADs using different methods. 
First-order anisotropies describe how directed or diffuse a particle event is when measured at a spacecraft. In the case of strong particle scattering, the imprints of the source processes may have vanished due to transport effects. If, however, large anisotropies are still observed, this is interpreted as a good magnetic connection to the source and less dominant transport effects allowing more direct comparisons with source characteristics as for example observed remotely. 
Characterizing the anisotropies is therefore important before one connects the {SEP} in-situ parameters with solar eruption observations (see also Sect.~\ref{sec:use_cases_magn}).

Anisotropies are determined from  {SEP} PADs. The pitch angle of each single particle, that is the angle between the particle's velocity vector and the magnetic field vector, is not directly measured. Instead, the pitch angle of the center of the telescope viewing direction and its angular width are used to determine the pitch-angle ranges in space, in which particles are observed. Determining the measured PADs, which describes the particle intensity distribution in pitch-angle space, requires the coupling of energetic particle and magnetic field data sets. Then, one can calculate the first and, if enough viewing directions are available, second order anisotropies that characterize the directionality of the particle beam with respect to the magnetic field, and infer the degree of interplanetary scattering and diffusion the particles experienced. Most of the instruments available in the SOLER tools provide only four different viewing directions, which is why we use the weighted-sum method \citep{Bruedern2018} proposed for four-sector measurements to calculate the first-order anisotropy $A_1$:
\begin{equation}
    A_1 = 3\frac{\sum_{i=1}^{N} \delta \mu_i \, \mu_i \, I(\mu_i)}{\sum_{i=1}^{N} \delta \mu_i \, I(\mu_i)} = 3\frac{\sum_{i=1}^{4} \delta \mu_i \, \mu_i \, I(\mu_i)}{\sum_{i=1}^{4} \delta \mu_i \, I(\mu_i)}.\label{eq:anisotropy}
\end{equation} 
Here $\mu_i$ is the central pitch-angle cosine of the $i$th telescope, $\delta \mu_i$ is the pitch-angle cosine range of the telescope opening cone, and $I(\mu_i)$ is the observed particle intensity in the $i$th telescope. 

Figure~\ref{fig:sep_pad} shows an example of a solar energetic electron event observed by the four viewing directions of  {SolO}  {EPD}/ {EPT} (3rd panel from top) and the corresponding first-order anisotropy (black trace in the bottom panel). 
One of the main factors affecting the magnitude of the anisotropy is the pre-event background, which should be subtracted to reveal the unbiased anisotropy of an event. Furthermore, in cases of decaying backgrounds such as in the event shown in Fig.~\ref{fig:sep_pad}, subtraction of the background intensity should take into account the temporal dependence. An important part of the {SEP}-{PAD}-and-Anisotropy tool is therefore the background removal. The user defines a background interval (gray shaded region in Fig.~\ref{fig:sep_pad}) from which a potentially time-dependent trend is determined automatically. This is done by fitting a constant and a time-dependent exponential model, from which the better model is chosen based on the fits' reduced $\chi^2$. As the telescope gathers observations at different pitch-angles, the tool also determines whether the background is better modelled with pitch-angle dependent models. Finally, the background model is extrapolated forward in time (dashed line in the 3rd panel from top) and used for background subtraction. The 3rd panel below shows the background subtracted intensities, and the bottom panel shows the 1st-order background-subtracted anisotropy in magenta. Notably, the background removal significantly affects the determined anisotropy, showing that without a proper background subtraction  {SEP} event anisotropies are often underestimated. We note, that due to the extrapolation of the background fits, the uncertainties of this component increase with time and that it is up to the user to decide how long in time to trust the determined background-subtracted anisotropies. However, usually significant anisotropies are only observed during the early phases of  {SEP} events, which is still close in time to the background window and therefore should be most trustable. 

A further improvement of currently available methodology, supported by the tool, is the determination of uncertainties of the first-order anisotropy. Therefore, a combination of bootstrapping, considering Poisson errors of the observed  {SEP} counting rates \cite[see also][]{Ehlert2022}, and uncertainties resulting from the background fits are used. The uncertainty ranges are shown as light magenta bands around the first-order anisotropy lines (Fig.~\ref{fig:sep_pad}, bottom). Furthermore, the gray shades at the top and bottom of the anisotropy panel denote the maximum range of anisotropies, which can be determined with the current pitch-angle coverage (first panel of Fig.~\ref{fig:sep_pad}) of the instrument. 

The SEP-PAD-and-Anisotropy tool currently supports the SolO/EPD instruments EPT and  HET, Wind/3DP/SST and STEREO/SEPT. The tool provides visualization of the data and a step-by-step analysis to subtract the background and show the determined anisotropy in comparison with the non-background subtracted values.

\subsubsection{SEP energy spectra calculation and fitting}
To analyze SEP spectra, the SOLER project provides two tools: the \textbf{SEP-Spectra} tool and the \textbf{SEP-Fit-Spectra} tool. 

\begin{figure}
    \centering
    \includegraphics[width=0.53\linewidth]{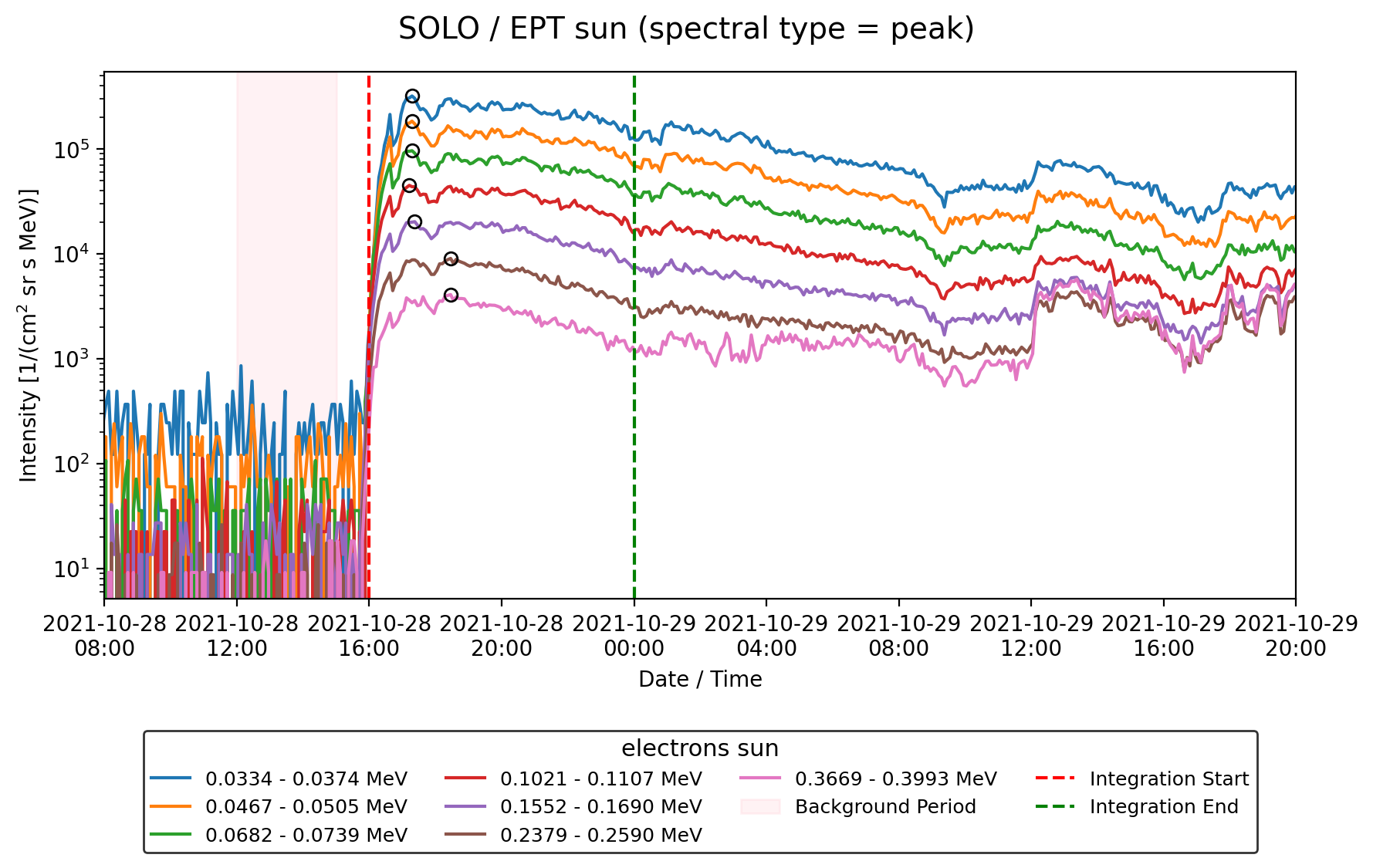}
    \quad
    \includegraphics[width=0.43\linewidth]{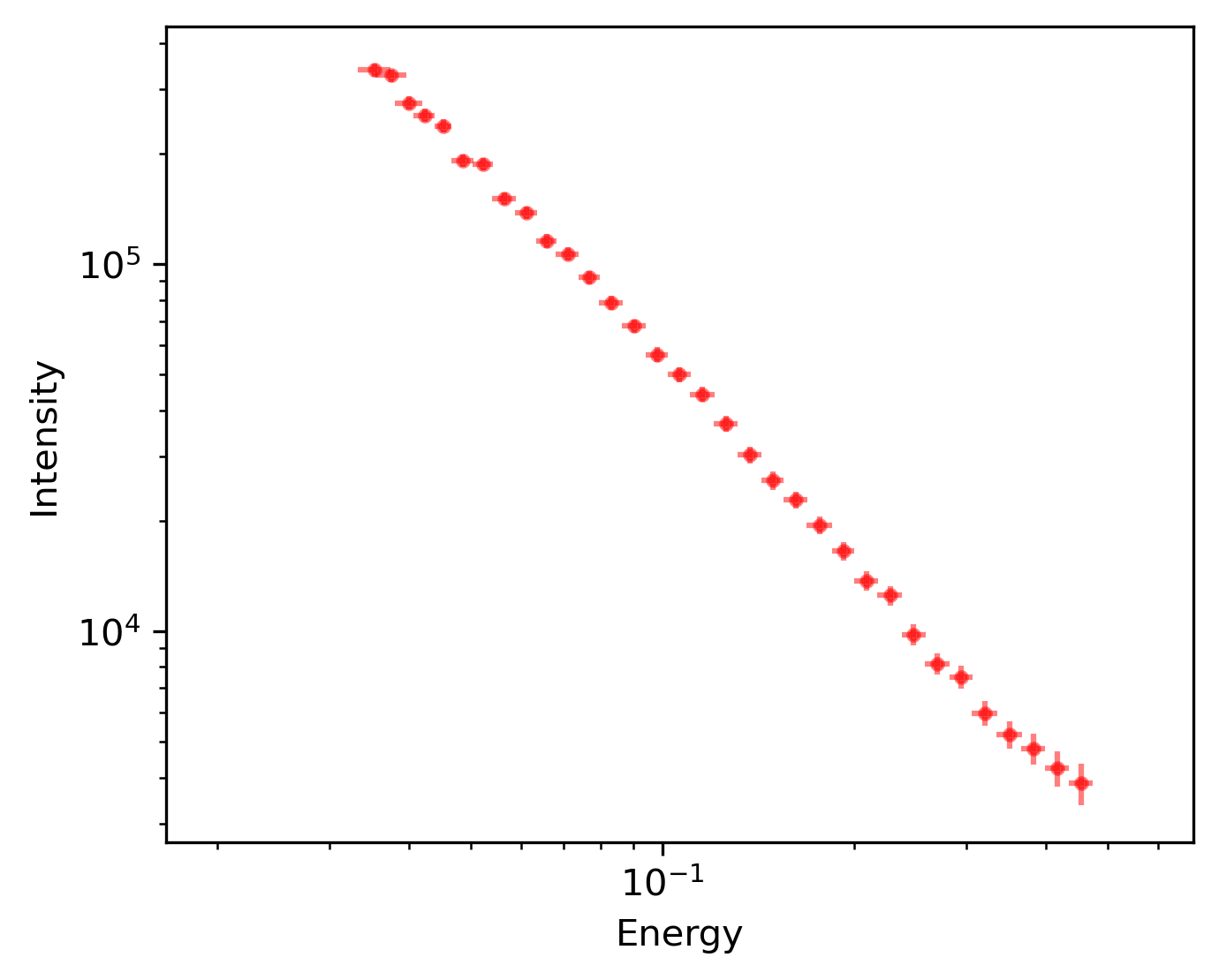}
    \caption{SEP-Spectra tool: Time series with selected background period (red shaded area), spectrum start and end times (dashed vertical lines), and determined peak intensities (black circles) on the left. The resulting background-subtracted peak spectrum is shown on the right.}
    \label{fig:SEP_spec_tool}
\end{figure}

The SEP-Spectra tool allows determining either peak-intensity spectra or accumulated spectra over a chosen time interval. A time series plot as shown in Fig.~\ref{fig:SEP_spec_tool} (left) supports the choice of time interval for the spectral determination as well as a background interval used for optional background subtraction. Then the spectrum is determined and a plot like Fig.~\ref{fig:SEP_spec_tool} (right) is displayed in the Notebook. A corresponding .csv file can be saved.
Furthermore, the tool allows visualizing the temporal evolution of the spectrum in an animated gif. For this purpose, the spectrum is determined for time slices of custom length within an overall time interval. The figures and corresponding .csv files for each time slice are automatically saved. 

\begin{figure}
    \centering
    \includegraphics[width=0.7\linewidth]{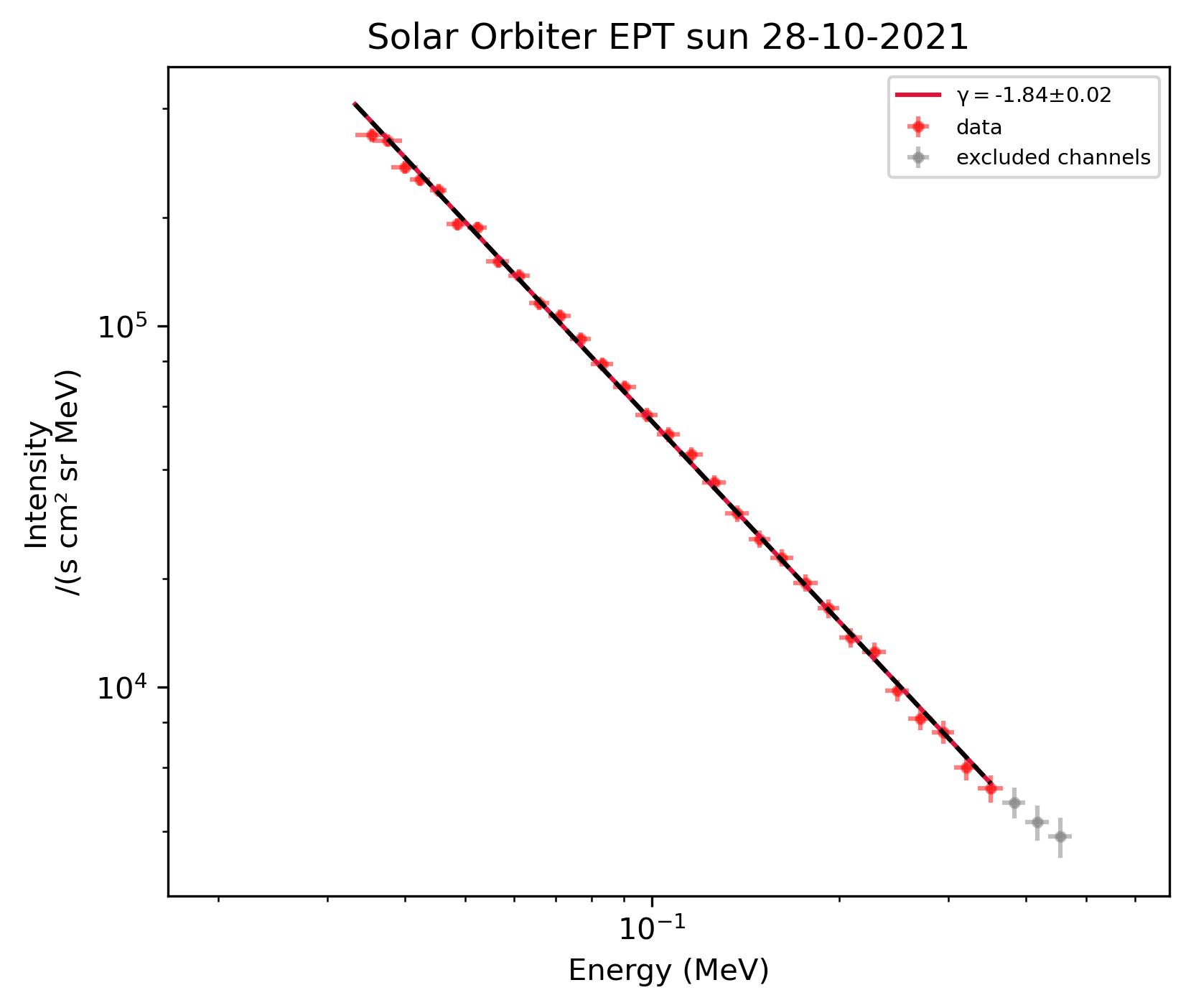}
    \caption{Example output of the {SEP}-Fit-Spectra tool showing an integral electron spectrum observed by {SolO} and fitted with a double-power-law function. The most important fit parameters and their uncertainties are provided in the legend.}
    \label{fig:spec_fit}
\end{figure}

The SEP-Fit-Spectra tool is directly compatible with the SEP-Spectra tool, as it ingests its .csv output files. However, the user can also load an own custom data file and fit the spectrum with the provided functionality. 
The fitting software used in the SEP-Fit-Spectra tool is described in \cite{Fedeli2026}. It currently includes five different power-law (pl) models: single  {pl}, double  {pl}, triple  {pl}, as well as the single- and double  {pl} models with an exponential cutoff at high-energies  \citep[see][and documentation in the Notebook for details]{Fedeli2026}. The  {SEP}-Fit-Spectra tool is able to fit the data with a specific model or to apply various models at the same time and chose the best fit from all models based on comparing their reduced $\chi^2$. The detailed fit results are then displayed in the Notebook (and can optionally be saved as a .csv file) along with a final output plot as shown in Fig.~\ref{fig:spec_fit}.

\section{Examples of analysis use cases}\label{sec:use_cases}

\subsection{Timing analysis}\label{sec:use_cases_timing}
\begin{figure}
    \centering
    \includegraphics[width=1\linewidth]{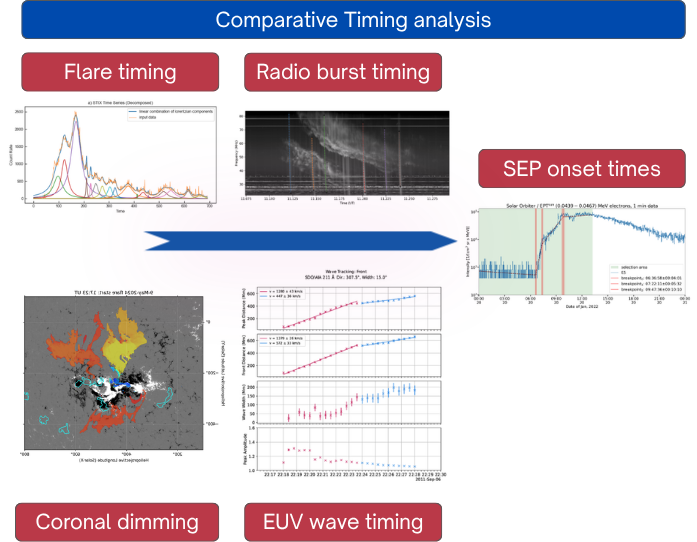}
    \caption{The times of various solar eruption features can be determined and compared using the SOLER tools. Starting at the Sun, the start of the flare and the times of various hard-X-ray peaks are determined with the \textit{SOLER X-ray timeseries analysis workflow}. \textit{SOLERsources} and \textit{SOLERwaves} further reveal times of flare dimming, and large-scale coronal wave times. Radio bursts timing is provided by \textit{NenuSunPy}. These times can then be compared with arrival times of SEPs at various observer positions using \textit{PyOnset} or the \textit{Onset-Regression} tool.}
    \label{fig:usecase_timing}
\end{figure}

In order to understand the causal connections of various solar eruption phenomena, a common starting point is to perform a comparative timing analysis. This is especially valuable to identify the parent solar source for SEPs observed in situ, especially in complex events that exhibit a zoo of different phenomena and potential acceleration regions and mechanisms.

Many of the SOLER tools can be used here to gain a comprehensive understanding on the timeline of an event allowing to identify connections and discrepancies between the various phenomena (see Fig.~\ref{fig:usecase_timing}). The start and peak time of the flare, including the time of various non-thermal X-ray peaks, can be extracted with the \textit{SOLER X-ray timeseries analysis workflow}. The start and end times of a coronal wave observed in EUV indicating when a coronal shock has formed, as well as when it passes certain regions in the low corona, are determined with the \textit{SOLER Wave Tool}. The \textit{Solersources} tool provides start times of coronal dimmings, which can be employed to infer the start of an erupting CME.
Start and end times of radio bursts as observed by NenuFAR are outputs of \textit{NenuSunPy}. They reveal when energetic particles are injected into interplanetary space as well as the presence of coronal and interplanetary shocks. The coronal modeling tools further support the timing analysis, for example, by revealing when a magnetic connection of a shock was established with a certain spacecraft (see also Sect.~\ref{sec:use_cases_magn} and \citet{morosan2025b}). Recent studies where type II radio burst positions intersect magnetic connections to spacecraft show that low energy and high energy electron injection times are close to the onset time of the type II bursts at the Sun \citep{morosan2024, morosan2025b}. However, a recent study where the spacecraft magnetic connections were farther from the flare site and location of type II bursts, show a delay of at least 20~minutes between high and low energy electrons and the onsets of type II bursts \citep{morosan2026}. Finally, SEP onset times can be determined with \textit{PyOnset} and the \textit{Onset-Regressionn Tool}. To better compare the SEPs with the solar phenomena one usually estimates their solar injection times by tracing the particles back to the Sun based on their speeds and the distance of the observer that largely determines their propagation path. The VDA functionality of PyOnset provides one way of estimating these SEP injection times.

\subsection{Magnetic connectivity analysis}\label{sec:use_cases_magn}
\begin{figure}
    \centering
    \includegraphics[width=1\linewidth]{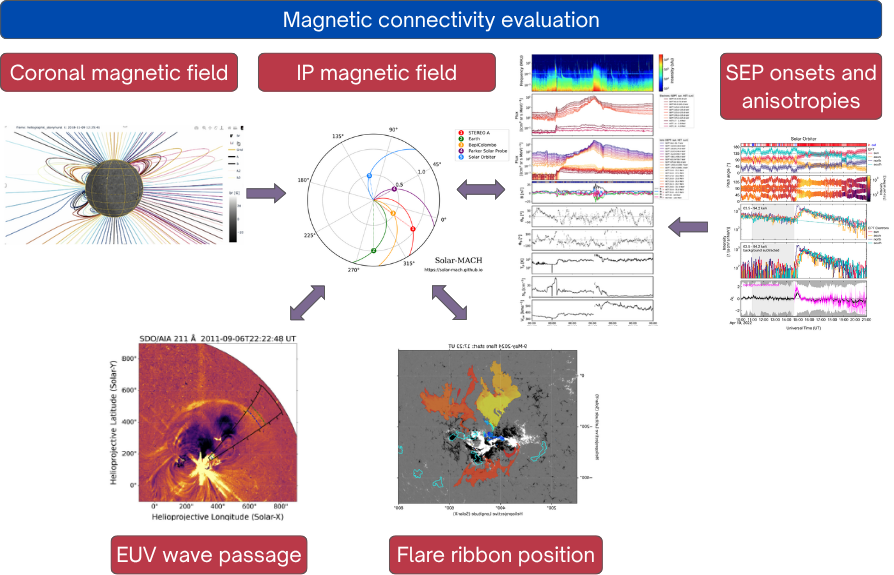}
    \caption{Comprehensive magnetic connectivity determination and testing using various SOLER tools. Coronal magnetic fields are provided by CIDER. Solar-MACH provides observer's magnetic footpoints at the Sun using a Parker spiral model and optionally a PFSS extension applied to regions close to the Sun. In-situ observations of SEP PADs and anisotropies as well as the in-situ magnetic field polarity can be used to probe the determined magnetic connection. Furthermore, in-situ magnetic field and plasma observations might reveal transient structures that caused deviations from the ideal Parker magnetic field. Detailed regions of interest for the magnetic connectivity analysis at the Sun are provided for example by the flare ribbon positions or regions passed by a large-scale coronal wave.}
    \label{fig:usecase_connectivity}
\end{figure}

To determine and probe the magnetic connection between solar activity regions at the Sun and interplanetary spacecraft that observe energetic particles related with a solar eruption, SOLER provides a variety of tools (see Fig.~\ref{fig:usecase_connectivity}). \textit{Solar-MACH} can be used to determine the magnetic connections in the interplanetary medium following an ideal Parker spiral model. Close to the Sun where the magnetic field often strongly deviates from the Parker model and becomes non radial, the PFSS extension of \textit{Solar-MACH }can be used. 
Similar functionality is also available in SHELVIS for determining the magnetic connectivity between the low and upper corona using other magnetic field models, in particular the steady-state magnetofrictional model as well as the Predictive Science MAS model. 
The determined magnetic connectivity can be furthermore probed by comparison with in-situ observations: The \textit{SEP Pitch-Angle and Anisotropies} tool can be used to test if the inferred magnetic polarity of the magnetic footpoint at the Sun fits the one observed in situ. This should be the case when the connecting field line is represented correctly by the models. Furthermore, if a direct magnetic connection to an SEP source region has been determined, it is expected that the SEP event shows significant anisotropy at the observer. Contrary, if no direct magnetic connection to the source is present, it is expected that the particles reach the observer through diffusive processes, for which less or completely vanished anisotropies are expected. It has to be noted, however, that the state of the interplanetary medium can significantly change magnetic connections as well as particle propagation conditions. Therefore, the \textit{Multi-Instrument} tool can be employed to investigate the local solar wind and magnetic field conditions around the time of the SEP event. This can help to identify interplanetary transient structures, which might have modified the determined magnetic connections. 

While we described above a workflow from an in-situ observer to the Sun, it is also possible to start from certain features at the Sun, for example, the regions passed by a large-scale coronal wave (\textit{SOLERwave} tool) or the location of the flare as for example determined by the \textit{SOLER X-ray imaging analysis workflow} or the \textit{SOLERsources} tool revealing the position of the flare ribbons. Magnetic connections between different coronal features and also interplanetary regions connecting to these can be traced with \textit{CIDER} and \textit{SHELVIS}.

\subsection{Energy spectra comparison}\label{sec:use_cases_spectra}

\begin{figure}
    \centering
    \includegraphics[width=1\linewidth]{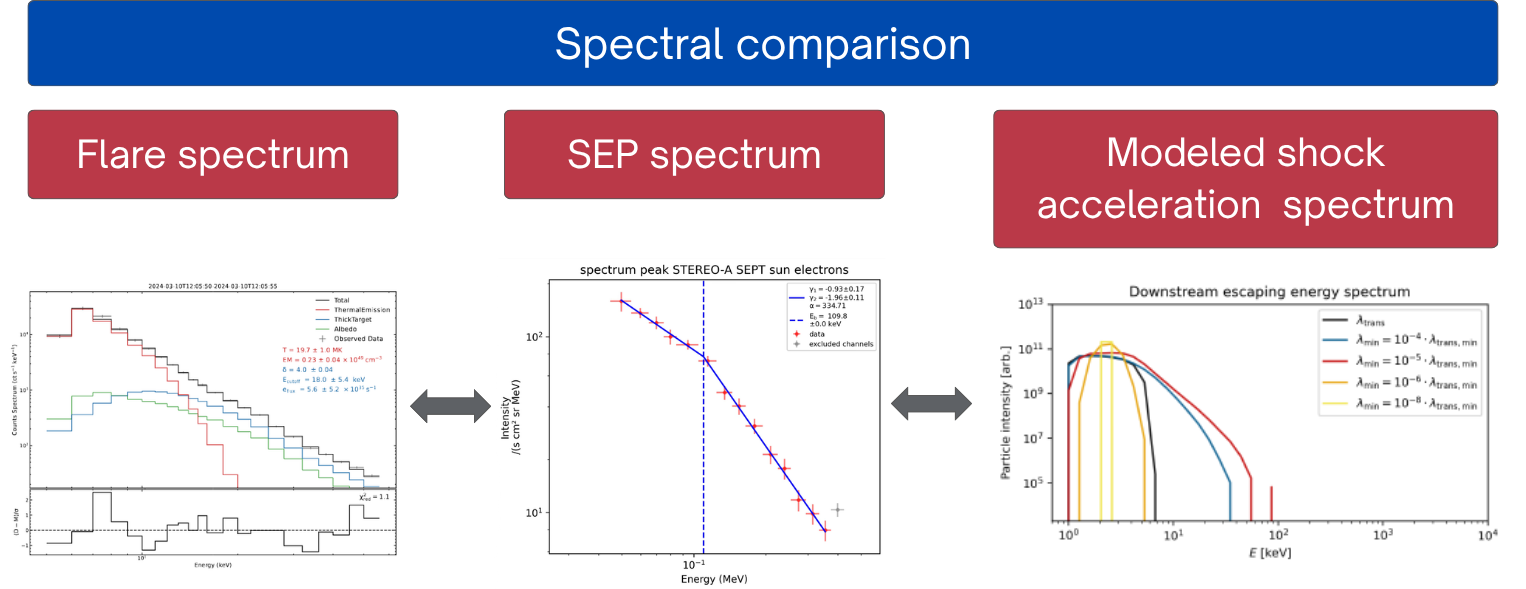}
    \caption{Use case illustration of comparative energy-spectra analyses. SEEs can either be accelerated within the solar flare or alternatively by a coronal shock. Comparing the energy spectra of SEEs with those of their potential sources can help identify the actual SEE source. Therefore, HXR spectra of the flare (left) are compared with the spectrum of electrons observed in situ at a spacecraft (center) for the same event. On the other hand, the SEE spectrum can be compared with energy spectra results of the electron-shock-acceleration model EASI (right).}
    \label{fig:usecase_spec}
\end{figure}

HXR spectra of the flare are assumed to constitute the same electron population like solar energetic electrons (SEEs) measured at a spacecraft in case of pure flare acceleration. We therefore expect their spectral indices to match \citep[e.g.,][]{Krucker2007, Dresing2021}. The SOLER tools enable such a spectral comparison by employing the \textit{SOLER X-ray spectroscopic analysis workflow} together with the \textit{SEP-Spectra} and \textit{SEP-Fit-Spectra} tools. Fig.~\ref{fig:usecase_spec} illustrates this by displaying the resulting HXR (left) and SEE spectra (center) including their fitted spectral parameter in the legends (see also Figs.~\ref{fig:xray_spec} and \ref{fig:spec_fit}. However, SEE spectra have also been linked to acceleration or re-acceleration by coronal shocks \citep[e.g.,][]{Krucker2007, Jebaraj2024}. One potential shock acceleration mechanism capable of accelerating electrons to high energies is Stochastic-Shock-Drift Acceleration \citep[SSDA][]{Amano2019}. The EASI model of SOLER (see Sect.~\ref{sec:easi}) applies this mechanism along with other mechanisms like Diffusive Shock Acceleration (DSA) and creates modeled SEE energy spectra (see Fig.~\ref{fig:usecase_spec} (right)) that can be then as well compared with the observed SEE spectrum. Based similarities between the different spectra and on which energy spectrum is closer to the observed SEE spectrum, conclusions can be drawn on the underlying acceleration process of the observed SEEs.

\section{Summary and conclusions}

We presented the SOLER project’s open analysis infrastructure for comprehensive, multi-messenger studies of energetic solar eruptions. SOLER combines three interlinked event catalogues of strong flares, fast CMEs, and energetic SEP events with a growing suite of openly available, Python-based analysis and modelling tools that span the full chain from solar source diagnostics to in-situ particle observations. Most of the tools are provided with Jupyter Notebook workflows, including application examples and analysis guidance, which lowers the barrier of getting familiar with a new tool and performing interdisciplinary analyses.

The currently released SOLER tools cover complementary observational regimes -- X-ray flare time series, spectroscopy and imaging; EUV/UV/H$\alpha$ feature detection of flare ribbons, dimmings, filaments; large-scale coronal wave tracking; radio dynamic-spectrum processing and imaging; coronal/shock-related modelling capabilities; magnetic connectivity assessment; and a broad set of SEP analysis methods including multi-spacecraft comparisons, onset determination, PAD and anisotropy analysis, longitudinal distribution characterisation, and energy spectra determination and fitting. Together, these capabilities enable coherent event timelines, improved Sun-to-spacecraft context for SEP interpretation, and quantitative cross-comparisons between flare, shock and particle signatures.

The SOLER tools are distributed as open-source software through the project’s GitHub community and most of them can also be executed without local installation via the SOLER JupyterHub, which provides maintained Python environments and immediate access through a web browser. This deployment model supports both rapid event studies and method development, and it enables users with different expertise levels to combine multiple diagnostic approaches within a single, consistent workflow.

Future releases will extend the tool set and data products, refine interoperability between the different analysis regimes, and further strengthen studies of solar eruptions across the expanding heliospheric spacecraft fleet. By providing openly accessible catalogues, data products, and analysis workflows, SOLER establishes a sustainable foundation for integrated investigations of solar eruption physics and for community-driven space-weather research.

\begin{acknowledgements}
The authors wish to acknowledge CSC – IT Center for Science, Finland, for computational resources.
Solar Orbiter is a mission of international cooperation
between ESA and NASA, operated by ESA. The STIX instrument is an international collaboration between Switzerland, Poland, France, Czech Republic, Germany, Austria, Ireland, and Italy.
\end{acknowledgements}

\begin{funding}
      This study has received funding from the European Union’s Horizon Europe research and innovation programme under grant agreement No.\ 101134999 (SOLER). This research reflects only the authors' view and the European Commission is not responsible for any use that may be made of the information it contains. 
      Work in the University of Turku and University of Helsinki was performed under the umbrella of Finnish Centre of Excellence in Research of Sustainable Space (FORESAIL) and Space Resilience funded by the Research Council of Finland (grant No.\ 352847 and No.\ 374097). 
      N.D. acknowledges support by the Research Council of Finland (SHOCKSEE, grant No.\ 346902 and AIPAD, grant No.\ 368509).
D.E.M acknowledges the Research Council of Finland project `SolShocks' (grant number 354409).
S. T. was supported by the German Space Agency (DLR), grant number 50 OT 2304.
\end{funding}

\begin{conflictofinterest}
    The authors declare no Conflict of Interest.
\end{conflictofinterest}

\begin{dataavailability}
      This article has no associated data generated and/or analyzed.
\end{dataavailability}

\bibliography{references}
   

\end{document}